\documentclass[preprint,aip,pop]{revtex4-1} 
\usepackage [english] {babel}
\usepackage {graphicx}
\usepackage {amsmath}
\usepackage {amsfonts}
\usepackage {amssymb}
\usepackage {amsthm}
\usepackage{bm}
\usepackage{hyperref}
\usepackage{datetime}
\newcommand{\al}{\lambda }
\newcommand{\IM}{\operatorname{Im}}
\newcommand{\RE}{\operatorname{Re}}
\newcommand{\bhat}{\bm{\hat{b}}}
\newcommand{\ExB}{$E\times B$\ }
\newcommand{\reff}[1]{(\ref{#1})}
\newcommand{\GA}[1]{\langle #1	 \rangle}
\newcommand{\pfrac}[2]{\frac{\partial#1}{\partial#2}}
\newcommand{\np}{\nabla_{\perp}}

\newcommand{\curl}[1]{\nabla \times #1}
\newcommand{\zhat}{\bm{\hat{z}}}

\begin{document}
\preprint{}

\title{ExB mean flows in finite ion temperature plasmas}

\author{J. Madsen}
\email{jmad@fysik.dtu.dk}
\author{J.~Juul Rasmussen} 
\author{V.~Naulin}
\author{A.~H. Nielsen}

\affiliation{Department of Physics, Technical University of Denmark,
DK-2800 Kgs. Lyngby,
Denmark}

\date{\today}

\begin{abstract}  
The impact of ion pressure dynamics on \ExB mean flows is investigated. Using a simplified, two-dimensional, drift ordered fluid model in the thin-layer approximation, three stresses in addition to the Reynolds stress are shown to modify the \ExB mean flow. These additional terms in the stress tensor all require ion pressure fluctuations. Quasi-linear analysis show that these additional stresses are as important as the Reynolds stress, and hence must be taken into account in analysis of transport barriers in which sheared \ExB mean flows are key ingredients. 
\end{abstract}
\maketitle

\section{Introduction}
Sheared mean flows are necessary for the formation of transport barriers\cite{wagner_2007} in magnetically confined plasmas. Transport barriers are always accompanied by a sheared radial electric field $E_r$ and an associated \ExB mean flow\cite{wagner_2007}, which in combination with flows along the magnetic field quench cross-field turbulent transport through decorrelation of turbulent eddies\cite{Biglari1990,Burrell1992}. Several mechanisms capable of driving mean flows have been suggested\cite{Connor2000}, but it is unclear whether the observed mean flows are due to a single motive force or whether they are a result of an interplay between many mechanisms. 

A particular mechanism for mean flow generation relies on the Reynolds stress tensor\cite{Reynolds1895}. It couples fluctuations and mean flows and hence renders  turbulence driven mean flows possible. In order to distinguish turbulence driven mean flows from equilibrium flows, turbulence driven mean flows are often called zonal flows. Both types of mean flows can suppress turbulence. In the fluid description the Reynolds stress originates from the advection non-linearity in the fluid momentum equation. By separating the velocity field into mean and fluctuating parts: $\bm{u} = \GA{\bm{u}} + \tilde{\bm{u}}$ and averaging the momentum equation one gets for an incompressible flow $\nabla \cdot \bm{u} = 0$:
\begin{align}
	\pfrac{\GA{\bm{u}}}{t} 
	+ \nabla \cdot \GA{\tilde{\bm{u}} \tilde{\bm{u}}}
	+  \nabla \cdot (\GA{\bm{u}} \GA{\bm{u}})
	= \mathcal{L},
\end{align}
where $\mathcal{L}$ represents forces, sinks, and sources. The average operation $\GA{\cdot}$ is unspecified here but is usually either a time-average, a flux surface average, or both. The Reynolds stress tensor $\GA{\tilde{\bm{u}} \tilde{\bm{u}}}$ can inhibit as well as enhance mean flows, but in strongly magnetized plasmas the approximate two-dimensional character of turbulence implies that energy is preferably transfered from smaller to larger scales\cite{Diamond1991, Scott2005NJP,Garcia2003PRE}. The energy transfer is between the kinetic energy of fluctuations and the kinetic energy of the mean flow. Therefore, Reynolds stress driven mean flows do not directly tap free energy but relies on conversion of free energy into fluctuating energy by other mechanisms\cite{Scott200353PLA}. On closed magnetic surfaces in strongly magnetized fusion plasmas, the mean convective term $\nabla \cdot (\GA{\bm{u}} \GA{\bm{u}})$ is usually negligible because gradients of the mean flow are to a good approximation perpendicular to the mean flow itself. 

When a plasma is subject to a strong confining magnetic field the dynamics is strongly anisotropic. Charged particles are approximately trapped on magnetic field lines along which they flow unhindered. When studying mean flows it is therefore convenient to apply models where this anisotropy is exploited a priori. The strong confining magnetic field implies that the magnetic dipole moment associated with the Larmor orbits of charged particles around magnetic field lines is an adiabatic invariant\cite{Alfven1940}. The invariance can be used in a dynamical reduction of the governing equations which lowers the computational costs by orders of magnitudes\cite{brizard_Hahm_review2007}. This is exploited in turbulence models which normally only consider dynamics on time scales longer than the inverse ion gyrofrequency\cite{brizard_Hahm_review2007, knorr_1988, Hinton_Horton_1971}. In the resulting equations the strong anisotropy imposed by the strong magnetic field appears explicitly. Velocities are split into perpendicular and parallel parts. In the direction perpendicular to the magnetic field advection is in most cases dominated by the \ExB-drift: $\bm{u}_E = \bm{E} \times \bm{B}/B^2$. Advection by other perpendicular fluid drifts associated with particle drifts such as the grad-B, curvature, and polarization drifts are inferior in comparison to the \ExB advection, but they are essential for the turbulence because the corresponding currents are dominant in the quasi-neutrality constraint $\nabla \cdot \bm{J} = 0$. In drift fluid models, which are used in this paper, the grad-B and curvature drifts and the magnetization current are contained in the diamagnetic drift $\bm{u}_D$\cite{Garcia_diamagnetic_fluid_part}. As in gyrokinetic\cite{brizard_Hahm_review2007} and gyrofluid models\cite{knorr_1988}, the diamagnetic and \ExB drifts are assumed to be of the same order of magnitude. However, since advection of all fluid fields by the diamagnetic drift cancels in all moment equations\cite{Tsai1970}, the diamagnetic flow is not responsible for transport over macroscopic distances. Therefore, it is only the mean \ExB flow  which is relevant in studies of decorrelation of turbulent eddies by perpendicular mean flows. 

In this paper we investigate how ion pressure dynamics influences \ExB mean flows. Reynolds stress driven mean flows have been studied extensively\cite{Diamond2005} and studies including ion pressure dynamics are numerous\cite{pop2003BDS,Scott2005NJP,Dif-Pradalier2009,Parra2010,Scott2010,Brizard2011,Krommes2013}. A common feature of these studies is that they do not consider "pure" mean flows but rather mean flows with multiple components. In gyrokinetic and gyrofluid treatments\cite{Dif-Pradalier2009,Parra2010,Scott2010,Brizard2011, Krommes2013}, the results concern mean flows, actually mean gyro-center momentum densities, in gyro-center coordinate space. Gyro-center space is a mathematical construction which provides tractable equations describing the dynamics down to gyro-radius length scales. The use of gyro-center coordinates is motivated by the notorious tedious expressions\cite{Smolyakov1998,Krommes2013} associated with gyro-radius length scale dynamics entering models expressed in standard coordinates. However, gyro-center coordinates are by construction not only functions of position and velocity but also of the electromagnetic potentials. To illustrate this point we express the zeroth order gyro-center moment, the gyro-center density $N$, in terms of physical quantities such as the particle density $n$, the ion scalar pressure $p_i$, and the electric potential $\phi$. In a quasi-neutral plasma $n_i = n_e$ we get\cite{madsen2013GF, MadsenPPCF2015} 
\begin{align}
	N_i = n_i - 
	\nabla_{\perp}^2 \bigg(\frac{p_{i} }{2m_i\Omega_i^2}\bigg) - \nabla \cdot \bigg( \frac{n_i}{B\Omega_i} \nabla_{\perp} \phi\bigg) 
\end{align}
where only terms to second order in $k_{\perp} \rho_i$ are retained. Here,  $k_{\perp}$ is a characteristic inverse gradient length scale,$\rho_i$ is the ion gyro-radius, $p_i$ is the ion pressure, and $\Omega_i = q_i B/m_i$ is the ion gyro frequency, where $q_i$ and $m_i$ are the ion charge and mass, respectively. The perpendicular projection of the gradient operator is defined as $\nabla_{\perp} = -\bhat \times (\bhat \times \nabla)$, where $\bhat = \bm{B} /B$ is a unit vector parallel to the magnetic field $\bm{B}$. Results formulated in gyro-center coordinates are therefore only directly relevant for the dynamics of gyrocenters, which is of course highly relevant, but in order to translate these results to measureable quantities the results must be transformed to well-known physical variables, a process which is tedious\cite{scott:102318,Krommes2013}. In low-frequency fluid models\cite{Hinton_Horton_1971} another but related issue appears. Here, the dominant perpendicular drifts are the fluid \ExB  and diamagnetic velocity fields. In previous works \cite{pop2003BDS,Scott2005NJP,Ramos2005,Parra2010} only the momentum and mean flow equations for the combined \ExB and ion diamagnetic flow were considered. This approach is problematic because the mean flow then includes the diamagnetic flow, which is not responsible for transport on the macroscopic length scale. 

The main objective of this paper is to investigate the \ExB mean flow and hence to disentangle the \ExB and ion diamagnetic parts. Considering the pure \ExB mean flow significantly complicates the governing equations. We have therefore deliberately chosen a paradigmatic, electrostatic drift fluid model in two-dimensional slab geometry, where dynamics along the magnetic field has been omitted. The model is presented in Sec.~\ref{sec:Model}. Even in this simplistic setup we show in Sec.~\ref{sec:meanflow} that the \ExB mean flow can be modified by four terms: i) The pure \ExB Reynolds stress $\GA{\tilde{\bm{u}}_E\tilde{\bm{u}}_E}$ and ii) a diamagnetic Reynolds stress\cite{Scott2005NJP} proportional to $\GA{u_y \partial_y p_i}$, where the $u_y$ denotes the "azimuthal" component of the \ExB drift. iii) We also show that \ExB mean flows may be driven by a term proportional to $\GA{\xi p_i u_x}$ in the stress tensor which is only finite when the magnetic field is inhomogeneous $\xi = 1/R \neq 0$, where $R$ is the major radius. iv) Lastly we demonstrate the existence of a component proportional to $2/3\GA{\xi p_i \partial_y p_e}$ of the stress tensor, which does not require \ExB drift fluctuations. The corresponding energy transfer terms, also commonly denoted production terms, are analyzed and conditions for enhancement and attenuation of \ExB mean flows for the individual energy transfer channels are determined. Next, in Sec.~\ref{sec:linearanalysis} we proceed with a quasi-linear analysis which reveals that that none of the four mean flow generation mechanisms are negligible. Lastly, our results are summarized and discussed in Sec.~\ref{sec:conclusion}.

\section{Model}\label{sec:Model}
This study uses an electrostatic drift fluid model\cite{Hinton_Horton_1971,Garcia_POP_062309, pop2003BDS,madsen2016} well-suited for studies of low-frequency turbulence in strongly magnetized plasmas particularly in the edge and scrape-off layer regions. 
The derivation of the model relies on the \textit{drift ordering} and hence on the existence of the small parameters:
\begin{align}
	\frac{\omega}{\Omega_i} \sim \epsilon \ll 1, \qquad
	\frac{u}{c_s} \sim \frac{\rho_s}{L_{\perp}}\sim\delta \sim \sqrt{\epsilon}, \qquad  
\frac{\rho_s}{L_B} \sim \epsilon_B \sim \delta^3.
	\label{eq:driftordering} 
\end{align}
That is, the model is only applicable to studies of low-frequency dynamics where the characteristic frequency $\omega$ is much smaller than the ion gyrofrequency $\Omega_i = q_iB/m_i$. Here, $B$ is the magnitude of the magnetic field, and $q_i$ and $m_i$ denote ion charge and mass, respectively. Further, the ordering presupposes that the fluid velocity $u$ is smaller than the ion sound speed $c_s = \sqrt{T_e/m_i}$, where $T_{e}$ is the electron temperature, and that the characteristic gradient length scale $L_{\perp}$ is longer than the hybrid ion gyroradius $\rho_s = c_s/\Omega_i$. Finally, the gradient length scale $L_B$ of the magnetic field is described by the small parameter $\epsilon_B$.

An advantage of the drift ordering is that algebraic expressions for the perpendicular part of odd fluid moment equations can be derived by a perturbative expansion in the small parameters. For instance, in a simple quasi-neutral plasma ($n = n_e \simeq n_i$), the terms on the right hand side of the momentum equation 
\begin{align}
	n m_a (\partial_t + \bm{u}_a \cdot \nabla ) \bm{u}_a
	+ \nabla \cdot \pi_a 
	= -\nabla p_a 
	+q_a n (\bm{E} + \bm{u}_a \times \bm{B})
	\label{eq:momentumequation}
\end{align}
dominate and balance to lowest order under drift ordering. Here, the subscript $a$ is a species label, $p_a$ is the scalar pressure, $\bm{E} = -\nabla \phi$ is the electric field, $\phi$ is the electrostatic potential, $\bm{B}$ is the magnetic field, and $\pi_a$ denotes the gyroviscous tensor. Therefore, the zeroth order perpendicular drifts are given as:
\begin{align} 
	\bm{u}_{\perp,0a} &= \bm{u}_E + \bm{u}_{Da}
	= \frac{\bhat \times \nabla \phi}{B}
	+ \frac{\bhat \times \nabla p_a}{q_anB}. 
\end{align}
By expanding the perpendicular velocity in $\epsilon$, the first order drifts, that represent the small terms on the left hand side of  Eq.~\reff{eq:momentumequation}, become:
\begin{align}
	\bm{u}_{\perp,1} &= \bm{u}_{p} + \bm{u}_{\pi} 
	= \frac{1}{\Omega} \bhat\times \frac{d}{dt} \bm{u}
	+ \frac{\bhat \times \nabla \cdot \pi}{qnB}.
\end{align}
The zeroth order drifts are the familiar \ExB-drift $\bm{u}_E$ and the diamagnetic drift $\bm{u}_D$, and the first order drifts are the polarization drift $\bm{u}_p$ and a gyroviscous drift $\bm{u}_{\pi}$. Inertia is described by the polarization drift. The dominant effect of the gyroviscous drift is to cancel the advection of momentum by the diamagnetic drift. This cancellation is in the literature refered to as the \textit{gyro-viscous cancellation}\cite{Hinton_Horton_1971,Brizard_1992,Smolyakov1998,Belova_2001,madsen2016}.  The first order drifts in $\bm{u}_{\perp,1}$ depend on the species mass, and hence only the ion drifts are retained.

In this study we investigate the influence of ion pressure dynamics on the generation, sustainment, and damping of mean flows. For this purpose and for the convenience of exposition we neglect the time-evolution of the parallel momentum and consider only the drift ordered equations governing the time evolution of vorticity and electron and ion pressure\cite{Nielsen2015,rasmussen_EPS_2015,madsen2016}:  
\begin{subequations}
\begin{align}
	       \nabla\cdot (n\bm{u}_{p_i})
	       +	       \nabla\cdot (n\bm{u}_{\pi i})
    +\nabla \cdot \big( n (\bm{u}_{Di}-\bm{u}_{De})\big)
= \Lambda_w,\label{eq:df_w}\\
    \frac{3}{2}\pfrac{}{t}p_i
      + \frac{3}{2}\nabla \cdot \big(p_i[\bm{u}_E + \bm{u}_{Di} + \bm{u}_{p_i}+ \bm{u}_{\pi} ]\big)
      +p_i \nabla \cdot [ \bm{u}_E +\bm{u}_{Di}  + \bm{u}_{p_i}+ \bm{u}_{\pi}] 
      + \nabla_{\perp} \cdot \bm{q}_i^*
= \Lambda_{p_i},
\label{eq:df_pi}\\
    \frac{3}{2}\pfrac{}{t}p_e
      + \frac{3}{2}\nabla \cdot \big(p_e[\bm{u}_E + \bm{u}_{De}] \big)
      +p_e \nabla \cdot [ \bm{u}_E +\bm{u}_{De} ] 
      + \nabla_{\perp} \cdot \bm{q}_e^*
= \Lambda_{p_e},
\label{eq:df_pe}
\end{align} 
\end{subequations}
where the diamagnetic heat flux is given as
\begin{align}
	\bm{q}^*_a = \frac{5}{2} p_a\frac{\bhat \times \nabla T_a}{q_a B}.  
\end{align}
The terms $\Lambda_w, \Lambda_{p_i}$ and $\Lambda_{p_e}$ on the right hand sides of Eqs.~\reff{eq:df_w}-\reff{eq:df_pe} represent, unspecified, parallel dynamics, collisional effects, and sources and sinks. We restrict the model to a local 2D slab geometry $(x,y,z)$ at the outboard midplane with the unit vector $\zhat$ aligned with the inhomogeneous magnetic field $\bm{B} = B(x)\zhat$. Periodic boundary conditions are invoked in the $y$-direction. 

The vorticity equation \reff{eq:df_w} is derived from the quasi-neutrality constraint $\nabla \cdot \bm{J} = 0$ using the electron and ion continuity equations (not shown here).  The diamagnetic drift represents the grad-B and curvature drifts, and diamagnetism due to gyration, which do not contribute to any particle transport over macroscopic distances when the magnetic field is constant. Therefore, all terms in the vorticity equation are of order $\epsilon^2$ despite that the diamagnetic current is of order $\epsilon$. In the vorticity equation \reff{eq:df_w} we make the thin-layer approximation\cite{Scott1998,Zeiler, madsen2016}. The approximation neglects particle density variations in the polarization and gyroviscous fluxes in the vorticity equation. The approximation resembles the Boussinesq approximation\cite{kundu2010fluid} in neutral fluid dynamics and is commonly invoked but is only strictly valid in regions with small particle density variations. Explicitly, the polarization and gyroviscous fluxes in the vorticity equation are approximated as\cite{madsen2016} 
\begin{align}
	\nabla\cdot (n\bm{u}_{pi})
    +	       \nabla\cdot (n\bm{u}_{\pi_i})
    \simeq
    - \nabla \cdot \bigg[
    \frac{n_0}{\Omega_0} \bigg(\pfrac{}{t} + \frac{B}{B_0} \bm{u}_E \cdot \nabla \bigg) \bigg(\frac{\np \phi}{B_0} + \frac{\np p_i}{q_in_0 B_0}\bigg)  	       	\bigg]
    \label{eq:thinlayer_up}
\end{align}
where $n_0$, $B_0$ and $\Omega_0 = e B_0 /m_i$ are characteristic, constant values of the particle density, the magnetic field, and the ion gyrofrequency, respectively. Here, the magnetic field is taken constant everywhere for two reasons: first, under drift ordering the variation of the background magnetic field in our local domain is minute. Secondly, energy conservation in models making the thin-layer approximation requires that the magnetic field in the polarization and gyroviscous fluxes is kept constant\cite{Scott2005NJP,madsen2016}. The absence of advection by the diamagnetic drift in equation \reff{eq:thinlayer_up} is due to the gyro-viscous cancellation\cite{Hinton_Horton_1971,Brizard_1992,Smolyakov1998,Belova_2001,madsen2016}. By inspection of Eq.~\reff{eq:thinlayer_up} we also see that the vorticity equation in fact governs the time evolution of the magnetic-field-aligned components of the \ExB and ion diagmagnetic vorticities: $\bhat \cdot \curl \bm{u}_{\perp,0 i}\simeq \nabla_{\perp}^2 \phi/B_0 + \nabla_{\perp}^ 2 p_i/(q_in_0B_0)$.

In the vorticity equation \reff{eq:df_w} thermal energy can be transformed into kinetic energy and hence drive instabilities and electrostatic turbulence. All terms in the vorticity equation are of order $\epsilon^2$ including the energy transfer terms. The time evolution of thermal energy is described by the pressure equations \reff{eq:df_pi}-\reff{eq:df_pe} where the leading order terms are of order $\epsilon$, but where the energy transfer terms evidently are of order $\epsilon^2$. 

However, to conserve energy, energy transfer terms balancing their counterparts in the vorticity equation are evidently of second order and must be retained to guarantee energy conservation. Without energy conservation, instabilities and hence turbulence may grow indefinitely in the absence of collisional dissipation, which would give incorrect saturated states. Furthermore, turbulence driven mean flows rely on similar energy transfer mechanisms, which also require energy conservation for a correct description of energy exchange between e.g., electrostatic fluctuations and \ExB mean flows. Therefore, we retain all second order terms in the pressure equations required for energy conservation. The remaining terms of order $\epsilon^2$ in the pressure equations are neglected. A detailed description of the second order terms are found in appendix \ref{app:second_order_p_terms}. 

To conclude, these approximations leave us with a paradigmatic, energy conserving model describing turbulence, lowest order finite Larmor radius (FLR) effects, ion temperature dynamics, and \ExB mean flows:
\begin{subequations}
	\begin{align}
	\frac{n_0}{\Omega_0}\nabla \cdot \bigg(\bigg[\pfrac{}{t} + \frac{\zhat \times \nabla \phi}{B_0} \cdot \nabla \bigg] \bigg[ \frac{\np \phi} {B_0}  + \frac{\np p_i}{q_i B_0 n_0}\bigg]\bigg) 
	+ \nabla \cdot (n[\bm{u}_{De} - \bm{u}_{Di}])
	= \Lambda_w,\label{eq:w}
	\\
	\frac{3}{2}\bigg[\pfrac{}{t} + \frac{\zhat \times \nabla \phi}{B_0} \cdot \nabla \bigg]p_i
	+p_i \nabla \cdot \bm{u}_E
	+\frac{p_i}{n_0}\nabla \cdot (n[\bm{u}_{De} - \bm{u}_{Di}])
	= \Lambda_{p_i},           
	\label{eq:ionpressure}\\
	\frac{3}{2}\bigg[\pfrac{}{t} + \frac{\zhat \times \nabla \phi}{B_0} \cdot \nabla \bigg]p_e
	+p_e \nabla \cdot \bm{u}_E
	= \Lambda_{p_e},           
	\label{eq:elecpressure}
	\end{align}
\end{subequations}
where the compression of the polarization and gyroviscous drifts "$p_i \nabla \cdot (\bm{u}_{p_i}+ \bm{u}_{\pi_i})$" in the ion pressure equation \reff{eq:df_w} were eliminated using the vorticity equation \reff{eq:df_pi}. Contributions from $\Lambda_w$ in the ion pressure equation have been absorbed in the redefined $\Lambda_{p_i}$. Our model resembles other local drift fluid models (see e.g. Refs.~\onlinecite{Scott1998,Zeiler}), but in these models dependent variables are linearized e.g., $p_i \nabla \cdot \bm{u}_E \simeq p_{i0} \nabla \cdot \bm{u}_E$.

It is convenient to introduce the Gyro-Bohm normalization 
\begin{align}
	\Omega_{i0} t \rightarrow t, \quad
	\frac{x}{\rho_s} \rightarrow x, \quad
	\frac{p_{e,i}}{p_{e0}} \rightarrow p_{e,i},\quad
	\frac{e\phi}{T_{e0}} \rightarrow \phi, \quad
\label{eq:gyro_bohm_normalization}
\end{align}
which allows us to recast the model in the following simple form:
\begin{subequations}
 \begin{align}
 \nabla \cdot \big(\frac{d}{dt}  \np \phi^*\big) 
    +\xi\pfrac{}{y}(p_e + p_i)    
     &= \Lambda_w,\label{eq:w_norm}
     \\
\frac{3}{2}\frac{d}{dt}p_i
      - p_i \xi\pfrac{\phi}{y}
      + p_i  \xi \pfrac{}{y}(p_e + p_i)    
           &= \Lambda_{p_i},           
\label{eq:ionpressure_thinlayer}\\
\frac{3}{2}\frac{d}{dt}p_e
      - p_e \xi\pfrac{\phi}{y}
           &= \Lambda_{p_e},           
\label{eq:elecpressure_norm}
\end{align}
\end{subequations}
where $\xi = \frac{\rho_s}{R}$ is the curvature constant and $R\sim L_B$ denotes the major radius.  Note that the gyro-Bohm normalization was introduced to simplify algebraic manipulations in the subsequent sections, but does not capture the characteristic length and time scales of the model which are larger and longer typically\cite{jmad2011FLRBlob} of the order of $L_{\perp}$ and $c_s^{-1} \sqrt{L_{\perp} L_B}$, respectively. The advective derivatives are defined as 
\begin{align}
	\frac{d}{dt} = \pfrac{}{t} + \{\phi,\cdot\},
\end{align}
where the \ExB-advection is written in terms of the anti-symmetric bracket 
\begin{align}
	\{f,g\}  = \pfrac{f}{x}\pfrac{g}{y} - \pfrac{f}{y}\pfrac{g}{x},
\end{align}
and the modified potential is defined by
\begin{align}
    \phi^* = \phi + p_i.
\end{align}
It is notable that the particle density is absent from the model if we disregard collisions. This feature is mainly due to the thin-layer approximation.

\subsection{Energy theorem}\label{sec:energythm}
The conserved energy is derived in two steps. First, the electron and ion pressure equations \reff{eq:elecpressure_norm}-\reff{eq:ionpressure_thinlayer} are integrated neglecting surface terms. Next the vorticity equation \reff{eq:w_norm} is multiplied by "$- \phi$" and integrated again neglecting surface terms. Adding the results we get
\begin{align}
       \frac{d}{dt}\int d\bm{x} \, \mathcal{E}   
      =\int d\bm{x}\, S_{\|},
      \label{eq:energy_theorem}      
\end{align}
where the energy density is given by
\begin{align}
      \mathcal{E}=  \mathcal{E}_i+ \mathcal{E}_e +\mathcal{E}^*
      = 
       \frac{3}{2} [p_i + p_e]
      + \frac{|\np \phi^*|^2}{2} ,
\end{align}
and 
\begin{align}
\mathcal{S}_{\|} = \Lambda_{p_i} + \Lambda_{p_e} - \phi^* \Lambda_w.
\end{align}
The energy density consists of the ion and electron thermal energy densities $\mathcal{E}_i$ and $\mathcal{E}_e$, respectively, and the "drift energy" density $\mathcal{E}^*$. The absence of the particle density $n$ and the magnetic field in the drift energy is a consequence of the thin-layer approximation invoked in the vorticity equation \reff{eq:w_norm}. The drift energy is a function of the modified potential $\phi^*$ and can be understood as the energy associated with the \ExB and diamagnetic drifts, or alternatively as describing the FLR corrected \ExB kinetic energy and FLR corrections to the ion thermal energy\cite{scott:102318,jmad2011FLRBlob,wiesenberger2014}. The time-evolutions of the individual parts of the integrated energy densities are given as
\begin{align}
 \frac{d}{dt}E^* =
 \frac{d}{dt}\int d \bm{x} \,\mathcal{E}^* &=
			   \int d \bm{x} \, - \xi [p_i+p_e]\pfrac{\phi}{y}
			   + \xi p_i  \pfrac{p_e}{y}
			   -\phi^*\Lambda_w,\label{eq:d_dt_drift_energy}\\
 \frac{d}{dt}E_i =
  \frac{d}{dt}\int d \bm{x} \,\mathcal{E}_i &= 
 			      \int d \bm{x} \, \xi p_i \pfrac{\phi}{y}
 			      - \xi p_i \pfrac{p_e}{y}
 			      + \Lambda_{p_i},\label{eq:d_dt_ion_energy}\\
 \frac{d}{dt}E_e =
\frac{d}{dt}\int d \bm{x} \,\mathcal{E}_e &= 
			   \int d \bm{x} \, \xi p_e \pfrac{\phi}{y}
			   +\Lambda_{p_e}. \label{eq:d_dt_elec_energy}
\end{align}
There are two types of energy transfer channels: i) the finite compression of the \ExB drift\cite{Scott_2005}, represented by the $\xi p_i \partial_y\phi$ and $\xi p_e \partial_y\phi$ terms, allow an \textit{interchange} of thermal energy and kinetic energy. ii) The finite compression of the first order drifts are responsible for the second type of energy transfer channel. This effect is represented by the $\xi p_i \partial_y p_e$ terms.

\section{Mean flows}\label{sec:meanflow}
In this section we analyze how ion pressure dynamics influences \ExB mean flows in our two-dimensional interchange turbulence model presented in Sec.~\ref{sec:Model}. The analysis encompasses a derivation of a \ExB mean flow equation and an analysis of energy transport between free (thermal) energy, fluctuations and mean quantities.
 
In this paper the averaging operation defining mean quantities is a spatial average in the periodic $y$-direction direction
\begin{align} 
	\GA{f} = \frac{1}{L_y} \int_0^{L_y} dy\,  f. 
\end{align}
Here, $f$ is an arbitrary function and $L_y$ is the domain length in the $y$-direction. The fluctuating part is defined accordingly $\tilde{f} = f - \GA{f}$. Using the vorticity equation \reff{eq:w_norm} the time evolution\cite{pop2003BDS, Scott2005NJP} of the mean and fluctuating parts of the drift energy is obtained 
\begin{align}  
 \frac{d}{dt}  E_0^*= 
 \frac{d}{dt}  
 \int d\bm{x} \, \frac{1}{2} |\pfrac{\GA{\phi^*} }{x} |^2
    &= \int d\bm{x} \, - \pfrac{^2 \GA{\phi^*}}{x^2}\GA{ \pfrac{\tilde{\phi}}{y} \pfrac{\tilde{\phi}^*}{x}}
	  - \GA{\phi^*} \GA{\Lambda_w},
	  \label{eq:meangenenergy} \\
 \frac{d}{dt}  \tilde{E}^*= 
 \frac{d}{dt}	
    \int d\bm{x} \, \frac{1}{2} |\nabla_{\perp} \tilde{\phi}^* |^2
    &= \int d\bm{x} \,  \pfrac{^2 \GA{\phi^*}}{x^2} \GA{\pfrac{\tilde{\phi}}{y} \pfrac{\tilde{\phi}^*}{x}}
	  + \xi(p_e + p_i) \pfrac{\phi^*}{y} 
	  - \tilde{\phi}^* \Lambda_w. 
\label{eq:flucdriftenergy}
\end{align}
The time evolutions of the energy integrals given in Eqs.~\reff{eq:flucdriftenergy},\reff{eq:d_dt_ion_energy}, and \reff{eq:d_dt_elec_energy} reveal an energy transfer between $\tilde{E}^*$ and the ion and electron thermal energy densities $E_i$ and $E_e$  by the term: $\xi(p_e + p_i) \pfrac{\phi^*}{y}$. The first term on the right hand sides of both equations, the modified Reynolds stress production terms, yield a  energy transfer between  the mean and the fluctuating drift energies. This term includes the standard \ExB Reynolds stress production term $u_0'\GA{u_{x}u_y}$, where $u_x = -\partial_y \tilde{\phi}$ and $u_y = \partial_x \tilde{\phi}$ denote the x and y components of the fluctuating \ExB drift, respectively,  and $u_0'= \partial_x u_0$ is the shear of the mean \ExB flow 
\begin{align}
	u_0 = \pfrac{\GA{\phi}}{x}. 
\end{align}
The Reynolds stress production term describes an energy transfer due to fluctuating radial transport of azimuthal momentum in the presence of a sheared mean flow. However, due to the presence of the modified potential $\phi^*$ in the modified production term, it is also a function of the mean and fluctuating parts of the ion diamagnetic drift. Since no fields are advected by the diamagnetic drift, these extra terms lack an obvious interpretation. Furthermore, the interpretation of the drift energy density $\mathcal{E}^*$ itself is not immediately obvious. Since the particle density is advected by the \ExB drift, it is more informative to consider the time evolution of the integrated \ExB mean flow energy, the integrated fluctuating \ExB energy, and the residual drift energy defined as: 
\begin{align}
	E_0 = \int d\bm{x} \, \frac{u_0^2}{2}, \quad
	\tilde{E} = \int d\bm{x} \, \GA{\frac{|\nabla_{\perp} \tilde{\phi}|^2}{2}},\quad 
E_{\times} = \int d\bm{x} \, \GA{\frac{|\nabla_{\perp} p_i|^2}{2}} + \GA{\nabla \phi \cdot \np p_i}, 
\end{align}
respectively. The time-evolution of these energy integrals are derived from the vorticity equation \reff{eq:w_norm} and the ion pressure equation\reff{eq:ionpressure_thinlayer} 
\begin{align}
    \frac{d}{dt} E_0 &=
    \int d\bm{x} \,  
      \bigg[
    		\underset{\textbf{A}}{ \GA{u_y u_x}}
    		-\underset{\textbf{B}}{\GA{u_y \pfrac{p_i}{y} }}
    		-\underset{\textbf{C}}{\frac{2}{3}\xi \GA{p_i \pfrac{p_e}{y}} }
   		-\underset{\textbf{D}}{\frac{2}{3}\xi\GA{ p_i  u_x}}
    \bigg] u_0'
	  - \GA{\phi} \bigg[
	  		\underset{\textbf{E}}{\GA{\Lambda_w}    
	  		-\frac{2}{3} \pfrac{^2}{x^2}\GA{\Lambda_{p_i}}}
	  		\bigg],
	  		\label{eq:meanflow_energy}\\
    \frac{d}{dt}\tilde{	E}  &= 
     \int d\bm{x} \,       
    	\bigg[
    			-\underset{\textbf{A}}{ \GA{u_y u_x}} 
    			+\underset{\textbf{B}}{\GA{u_y \pfrac{p_i}{y}} }
    	\bigg]  u_0'
	  + \underset{\textbf{F}}{\xi\GA{(p_e + p_i) u_x} }
	  -\underset{\textbf{G}}{\frac{2}{3}\xi \GA{p_i  \np^2 \tilde{\phi} \pfrac{}{y}(p_i + p_e - \phi)}}\notag \\
	  &  \quad \quad - \GA{\tilde{\phi} \bigg[
	   \underset{\textbf{E}}{\Lambda_w	 
	   - \frac{2}{3}\pfrac{^2 }{x^2} \Lambda_{p_i} }
	   \bigg]},
	   \label{eq:fluc_energy}
	     \\
\frac{d}{dt} 	E_{\times} &= 
     \int d\bm{x} \,  
	   \underset{\textbf{H}}{\xi \GA{p_i \pfrac{p_e}{y}} }
	  +\bigg[
	  \underset{\textbf{C}} {\frac{2}{3} \xi \GA{p_i \pfrac{p_e}{y}} }
  	  +	  \underset{\textbf{D}}{ \frac{2}{3} \xi \GA{p_i u_x}}
  	  \bigg] u_0'
	  +	  \underset{\textbf{G}}{\frac{2}{3} \xi \GA{p_i\np^2 \tilde{\phi}   \pfrac{}{y}( p_e + p_i -\phi)}}\notag \\
	  & - \underset{\textbf{E}}{ \GA{p_i \Lambda_w}
	  - \frac{2}{3} \GA{\phi \pfrac{^2}{x^2}\Lambda_{p_i}}}.
	  \label{eq:crossterm_energy}
\end{align}
The energy integrals are accompanied by an equation for the mean \ExB flow, which is obtained by averaging the vorticity equation \reff{eq:w_norm} over the periodic $y$-direction making use of the ion pressure equation \reff{eq:ionpressure_thinlayer}  
\begin{align}
     \pfrac{u_0}{t} 
     + \underset{\textbf{a}}{\pfrac{}{x}\GA{u_xu_y}}
     -  \underset{\textbf{b}}{\pfrac{}{x}\GA{u_y\pfrac{p_i}{y}}}
     -\underset{\textbf{c}}{\frac{2}{3}\xi \pfrac{}{x}\GA{p_i \pfrac{}{y}p_e}} 
     -\underset{\textbf{d}}{\frac{2}{3}\xi \pfrac{}{x}\GA{p_i u_x}} 
     = \underset{\textbf{e}}{-\frac{2}{3} \pfrac{}{x} \GA{\Lambda_{p_i}}
	 + \int_0^x dx \, \GA{\Lambda_w}}   ,
     \label{eq:meanflow}
\end{align}
where boundary terms were neglected. Integrating the mean flow equation in the x-direction shows that no mean flow is generated without external sources. The time-evolution of the energy integrals and the mean flow equation are principal results of this paper. 

First, we note that the energy integrals and the mean flow equation reduce to the well-known system of equations in two-dimensional interchange driven convection\cite{garcia_inter} in the limit of constant ion pressure. Specifically, all ion pressure dependent terms vanish, $E_{\times} = 0$, and the time-evolution of the mean flow is governed by two effects: the divergence of the Reynolds stress tensor marked "\textbf{a}", which describes radial transport of azimuthal momentum, and collisional viscous damping marked "\textbf{e}". These two effects are accompanied by corresponding  energy transfer terms in the mean flow energy equation \ref{eq:meanflow_energy}. Collisional dissipation damps the mean flow energy through the term "\textbf{E}". The  Reynolds stress production terms marked "\textbf{A}" in equations \reff{eq:meanflow_energy} and \reff{eq:fluc_energy} yield a  energy transfer between the mean and fluctuating \ExB kinetic energies. From the energy integrals it is evident that the mean flow energy $E_0$ is only altered by the Reynolds stress when the mean flow is sheared $u_0'\neq 0$. The condition of a sheared mean flow is necessary but not sufficient. By expanding the electric potential into an infinite Fourier series in the periodic $y$-direction, the $x-y$ component of the Reynolds stress tensor can be written as 
\begin{align}
	\GA{u_x u_y}
	= - 2\sum_{k_y=1}^{\infty} k_y |\phi_{k_y}|^2 \delta'_{\phi}, 
	\label{eq:Reynolds_spectral}
\end{align}
where $|\phi_{k_y}(x,t)|$ and $\delta_{\phi}(x,t)$ denote the radially varying amplitude and phase, respectively, and $\delta'_{\phi} = \partial_x \delta_{\phi}$. The mean flow energy is therefore only altered if the mean flow is sheared \textit{and} if the phase of the electrostatic potential varies radially. The thermal and fluctuating energies are coupled through the term marked "\textbf{F}" whose origin is magnetic field inhomogeneity. This  energy transfer describes fluctuating radial transport of thermal energy. The spectral representation of this \textit{interchange drive term} is
\begin{align}
	   \xi\GA{p_e u_x}
	=  \xi\sum_{k_y=1}^{\infty} 2 k_y |\phi_{k_y}| |p_{ek_y}| \sin (\delta_{\phi} - \delta_{p_e})  	
	\label{eq:FETransSpec}
\end{align}
demonstrating that the direction of the energy flux is determined by the phase difference between electric potential and electron pressure fluctuations. Note that there is no direct energy transfer between the integral of the electron thermal energy  $E_e$ and the mean flow energy $E_0$; the only path for thermal energy to the mean flow energy goes through the fluctuating energy $\tilde{E}$. 

When the assumption of constant ion pressure is relaxed, additional mean flow sources emerge. First, the Reynolds stress in the mean flow equation \reff{eq:meanflow}, marked "\textbf{a}", is accompanied by a diamagnetic Reynolds-stress-like term, marked "\textbf{b}" and corresponding  production terms marked "\textbf{B}" in the mean and fluctuating energy integrals equations \ref{eq:meanflow_energy} and \ref{eq:fluc_energy}. Like the Reynolds stress production term, a finite energy transfer by the diamagnetic Reynolds energy transfer term requires a sheared mean flow $u_0' \neq 0$. The spectral representation in the $y$-direction
\begin{align}
	\GA{u_y \pfrac{p_i}{y} }
	=	\sum_{k_y>0} 2 k_y \bigg[
		\sin(\delta_{\phi} - \delta_{p_i}) |p_{ik_y}||\phi_{k_y}|' 
		+ \cos (\delta_{\phi} - \delta_{p_i}) |p_{ik_y}||\phi_{k_y}| \delta_{\phi}'  
	\bigg]
	\label{eq:DiaReynoldsStress}
\end{align}
shows that the diamagnetic Reynolds stress and the corresponding production term may modify the mean flow both when $\phi$ and $p_i$ are in and out of phase. Furthermore, the ability of the diamagnetic Reynolds stress production term to modify the mean flow does not require that the phase of the electric potential is radially inhomogeneous as is required for the standard Reynolds stress. We also note that if $\phi = - p_i + \text{const.}$, which is an approximate steady state solution to the vorticity equation \ref{eq:w_norm}, then the Reynolds and the diamagnetic Reynolds stresses cancel.  

\begin{figure}[hbtp]
\centering
\includegraphics[width=0.7\textwidth]{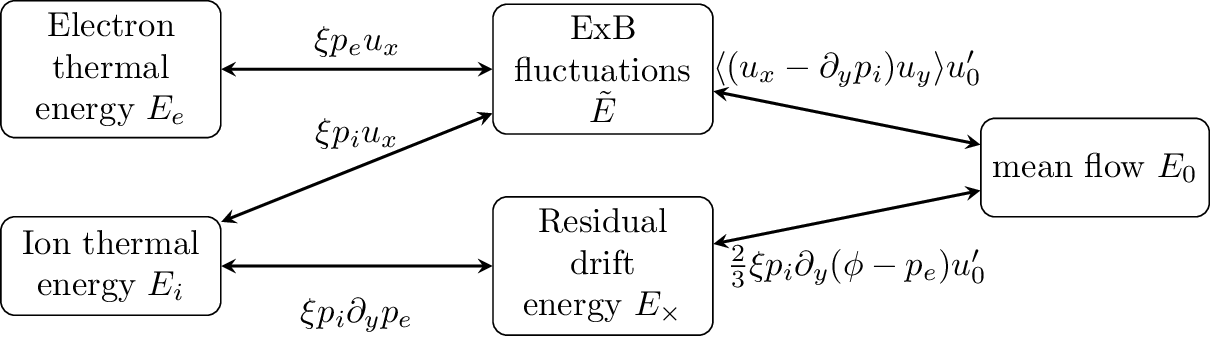}
 \caption{Diagram illustrating the energy transfer channels  between the five energy integrals in equations \reff{eq:d_dt_ion_energy}-\reff{eq:d_dt_elec_energy} and \reff{eq:meanflow_energy}-\reff{eq:crossterm_energy}. Energy transfer channels are shown as uni-directional arrows; the corresponding energy transfer terms label the arrows.} 
 \label{fig:energy_transfer}
\end{figure}

In addition to the diamagnetic Reynolds stress, two  transfer terms marked "\textbf{c}" and "\textbf{d}" enter the mean flow equation \reff{eq:meanflow} when the ion pressure is non-constant. These transfer terms differ from the standard and diamagnetic Reynolds stresses because of their ability to modify the mean flow rely on an inhomogeneous magnetic field $\xi \neq 0$. The corresponding energy transfer terms, marked "\textbf{C}" and "\textbf{D}" in equations \reff{eq:meanflow_energy} and \reff{eq:crossterm_energy}, couple  the mean flow energy $E_0$ and the residual energy $E_{\times}$. In the constant ion pressure limit, the fluctuating kinetic energy and therefore also instabilities can only grow because the fluctuations can feed on the thermal energy through the interchange drive term marked "\textbf{F}". When the ion pressure is not constant, an additional  energy transfer emerges. The term marked "\textbf{H}" in the residual energy integral equation \reff{eq:crossterm_energy} allows energy exchange between the residual energy and the ion thermal energy. In many respects the generation of mean flows in interchange driven turbulence is therefore potentially fundamentally different when ion temperature dynamics is taken into account. The energy transfer channels are schematically depicted in figure \ref{fig:energy_transfer}.
We note that the appearance of the terms "\textbf{C}","\textbf{D}", and "\textbf{H}" in the energy integral equations and the terms "\textbf{c}" and "\textbf{d}" in the mean flow equation is a direct consequence of consistently keeping the first order drifts in the ion density and in the ion pressure equations. The terms in equations \reff{eq:fluc_energy} and \reff{eq:crossterm_energy} marked "\textbf{G}" yield a  energy transfer between the fluctuating \ExB energy and the residual drift energy. We do not analyze these terms further in this paper. A detailed analysis most likely requires that the residual drift energy is split into mean and fluctuating components. We leave this analysis for future work. 

The term marked "\textbf{d}" in the mean flow equation \reff{eq:meanflow} originates from the finite compression of the \ExB drift in the ion pressure equation \ref{eq:ionpressure_thinlayer}. The spectral decomposition 
\begin{align}
	  \frac{2}{3} \xi\GA{p_i u_x}
	=  \frac{4}{3}\xi\sum_{k_y=1}^{\infty} k_y |\phi_{k_y}| |p_{ik_y}| \sin (\delta_{\phi} - \delta_{p_i})  	
	\label{eq:DETransSpec}
\end{align}
shows that a finite phase difference between the potential and ion pressure fluctuations is required for modification of the mean flow. It is interesting that this term apart from a factor "$2/3$" shares the same functional form as the interchange drive term "\textbf{F}" in the energy integral equation \reff{eq:crossterm_energy}, and hence they are always simultaneously active. The direction of the energy flux by the corresponding energy transfer terms marked "\textbf{D}" in Eqs.~\reff{eq:meanflow_energy} and \reff{eq:crossterm_energy} is determined by the phase shift and the mean flow shear. 

Finally, we analyze the transfer mechanisms described by the terms "\textbf{C}" and "\textbf{H}" in the energy integral equations \reff{eq:meanflow_energy} and \reff{eq:crossterm_energy} and the corresponding term "\textbf{c}"  in the mean flow equation \reff{eq:meanflow}. A remarkable feature of these terms is that they are independent of the fluctuating part of the \ExB drift, and hence may alter the mean flow when \ExB- drift fluctuations vanish $u_x=u_y= 0$. As illustrated in Fig.~\ref{fig:energy_transfer}, ion thermal energy  $\mathcal{E}_i$ can be transferred to the mean flow energy $\mathcal{E}_0$ via the residual energy $\mathcal{E}_{\times}$ by these transfer channels. Common to all these terms is the appearance of 
\begin{align}
	\xi\GA{p_i\pfrac{p_e}{y}}
	= - \xi\sum_{k_y=1}^{\infty}  2k_y |p_{ek_y}| |p_{ik_y}| \sin (\delta_{p_e}- \delta_{p_i}),
	\label{eq:pe_ddy_pi}
\end{align}
showing that they are only active if the phase shift between electron and ion pressure fluctuations is finite. It is important to keep in mind that these terms vanish in the isothermal limit; electron or ion temperature fluctuations are required. The direction of the energy flux through the transfer channel "\textbf{H}" between the ion thermal energy $\mathcal{E}_i$ and the residual drift energy $\mathcal{E}_{\times}$ is solely determined by the phase-shift $\delta_{p_e}- \delta_{p_i}$. Specifically, energy is transported from the ion thermal energy to the residual drift energy when $\sin(\delta_{p_e}- \delta_{p_i})< 0$, and is maximal when $\delta_{p_e}- \delta_{p_i} = -\pi/2$. For the residual drift energy to flow simultaneously from the residual drift energy $\mathcal{E}_{ \times}$ to the mean flow energy $\mathcal{E}_0$, the shearing rate $u_0'$, entering the transfer term $2/3 \xi u_0' \GA{p_i \partial_y p_e}$ marked "\textbf{C}" in equations \reff{eq:meanflow_energy} and \reff{eq:crossterm_energy}, must be negative $u_0' < 0$. The neglected higher order terms in the pressure equations \reff{eq:ionpressure_thinlayer} and \reff{eq:elecpressure_norm} yield additional terms in the \ExB mean flow energy equation which can be found in appendix \ref{sec:appendix}. 

Recall that the results presented in this section are derived using the simplified model given in Eqs.~\reff{eq:w_norm}-\reff{eq:elecpressure_norm}, where some higher order terms in the pressure equations were neglected (see appendix \ref{app:second_order_p_terms}). Before proceeding, we note that our results are not qualitatively altered if all higher order terms were retained. As shown in appendix \ref{sec:appendix}: the energy theorem derived in section \ref{sec:energythm} and the mean flow equation \reff{eq:meanflow} are not changed. Two coefficients in the \ExB mean flow energy equation \reff{eq:meanflow_energy} change form $2/3$ to $5/3$, and two additional small terms are added. Furthermore, an equation governing the particle density must be added to the model. All things considered, the simplified model provide the same results, permits a clear exposition, and significantly simplifies the algebra in the derivations.

\section{Linear analysis}\label{sec:linearanalysis}
In this section we investigate the additional terms, beyond the Reynolds stress and associated production term, in the mean flow and energy integral equations which arise when ion temperature dynamics is taken into account. The analysis is carried out by means of linear and quasi-linear analysis. This approach allows us to estimate under which conditions these additional terms are active and to some extend to estimate their magnitude and whether they act as to inhibit or enhance mean flows

\begin{figure}[hbtp]
 \centering
\includegraphics[width=0.49\textwidth]{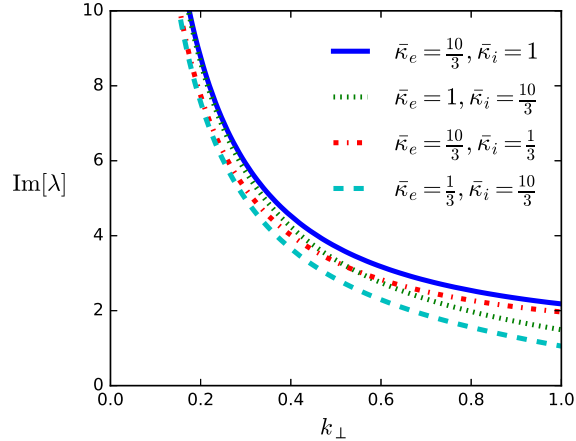}
 \caption{Growth rates for different values of the electron and ion inverse profile gradient length scales, $\bar{\kappa}_e$ and $\bar{\kappa}_i$, respectively. By comparing the red and green curves, we see the effect of ion FLR stabilization.} 
 \label{fig:DispRel}
\end{figure}

Neglecting dissipative effects assuming a local plane wave solution $\exp(i \bm{k}\cdot \bm{x} - i\omega t)$ to the model equations \reff{eq:w_norm}-\reff{eq:elecpressure_norm}, the linearized equations are
\begin{align}
	\omega k_{\perp}^2 (\phi_{\bm{k}} + p_{i\bm{k}}) + \xi k_y (p_{e\bm{k}} + p_{i\bm{k}}) = 0, \\
	-\frac{3}{2} \omega p_{i\bm{k}} + \phi_{\bm{k}} k_y (\frac{3}{2} \kappa_i - \xi) 
	+ \xi k_y (p_{e\bm{k}} +p_{i\bm{k}}) = 0, \\
	-\frac{3}{2} \omega p_{e\bm{k}} + \phi_{\bm{k}} k_y (\frac{3}{2} \kappa_e - \xi) 
	= 0,  	
\end{align}
with the dispersion relation 
\begin{align}
	\al \bigg[	\al^2
	+ \al (\bar{\kappa}_i - \frac{4}{3} )
	+  (\frac{2}{3}\bar{\kappa}_e  - \frac{4}{9} )
	+\frac{1}{k_{\perp}^2} (\bar{\kappa}_e  + \bar{\kappa}_i - \frac{4}{3})\bigg]
	= 0, 
\end{align}
where $\al = \frac{\omega}{\xi k_y}$, $\bar{\kappa}_i = \kappa_i/\xi$, $\bar{\kappa}_e=\kappa_e/\xi$, and $\kappa_i$ and $\kappa_e$   denote the ion and electron inverse profile gradient length scales, respectively. 
Besides the trivial solution $\al = 0 $, the dispersion relation has the solutions
\begin{align}
\al
= \frac{ \frac{4}{3} - \bar{\kappa}_i
\pm\sqrt{(\bar{\kappa}_i-\frac{4}{3})^2
-4 
k_{\perp}^{-2}(\bar{\kappa}_i+\bar{\kappa}_e-\frac{4}{3})}}{2}.
\label{eq:sol_disp_rel} 
\end{align}
The unstable part of the solution for which: $\IM(\lambda) > 0$, is plotted in Fig.~\ref{fig:DispRel} for various parameters.
Instability requires that $\bar{\kappa}_i + \bar{\kappa}_e > 4/3$.  Notice the well-known ion FLR stabilization\cite{weiland1999collective} by the first term in the radicand in Eq.~\reff{eq:sol_disp_rel}. The stabilizing effect is clearly illustrated by the blue and green curves in Fig.~\ref{fig:DispRel} which have the same interchange drive "$\bar{\kappa_i} + \bar{\kappa_e}$" but when $\bar{\kappa_e}> \bar{\kappa_i}$ (blue) the growth rate is significantly higher than when $\bar{\kappa_e}< \bar{\kappa_i}$ (green). Only for very low $k_{\perp}$ (not visible in Fig.~\ref{fig:DispRel}) the growth rate of the green curve exceeds the blue curve.

\begin{figure}[hbtp]
 \centering
\begin{minipage}[c]{.32\textwidth}
  \par\vspace{0pt}
  \centering 
\includegraphics[width=1\textwidth]{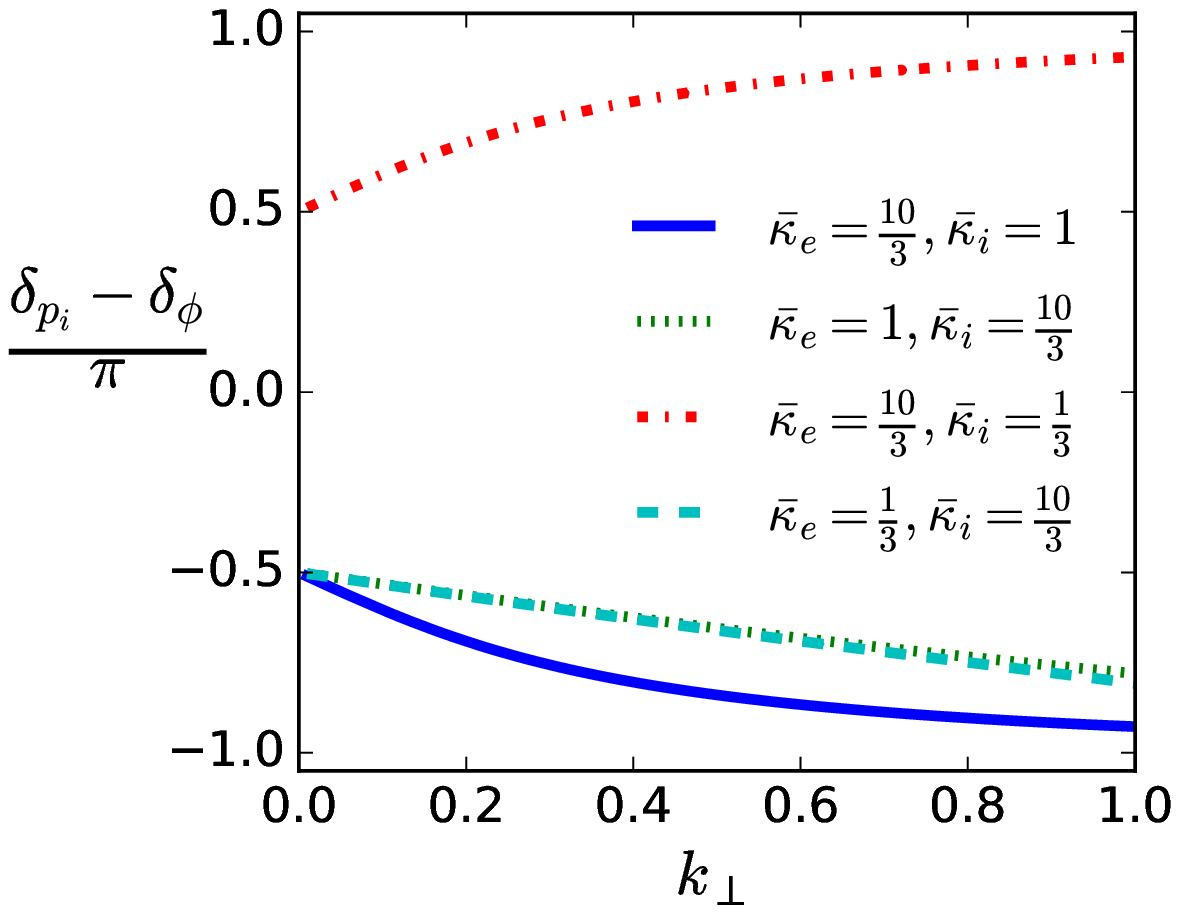}
 \end{minipage}
\begin{minipage}[c]{.32\textwidth}
  \par\vspace{0pt}
  \centering 
\includegraphics[width=1\textwidth]{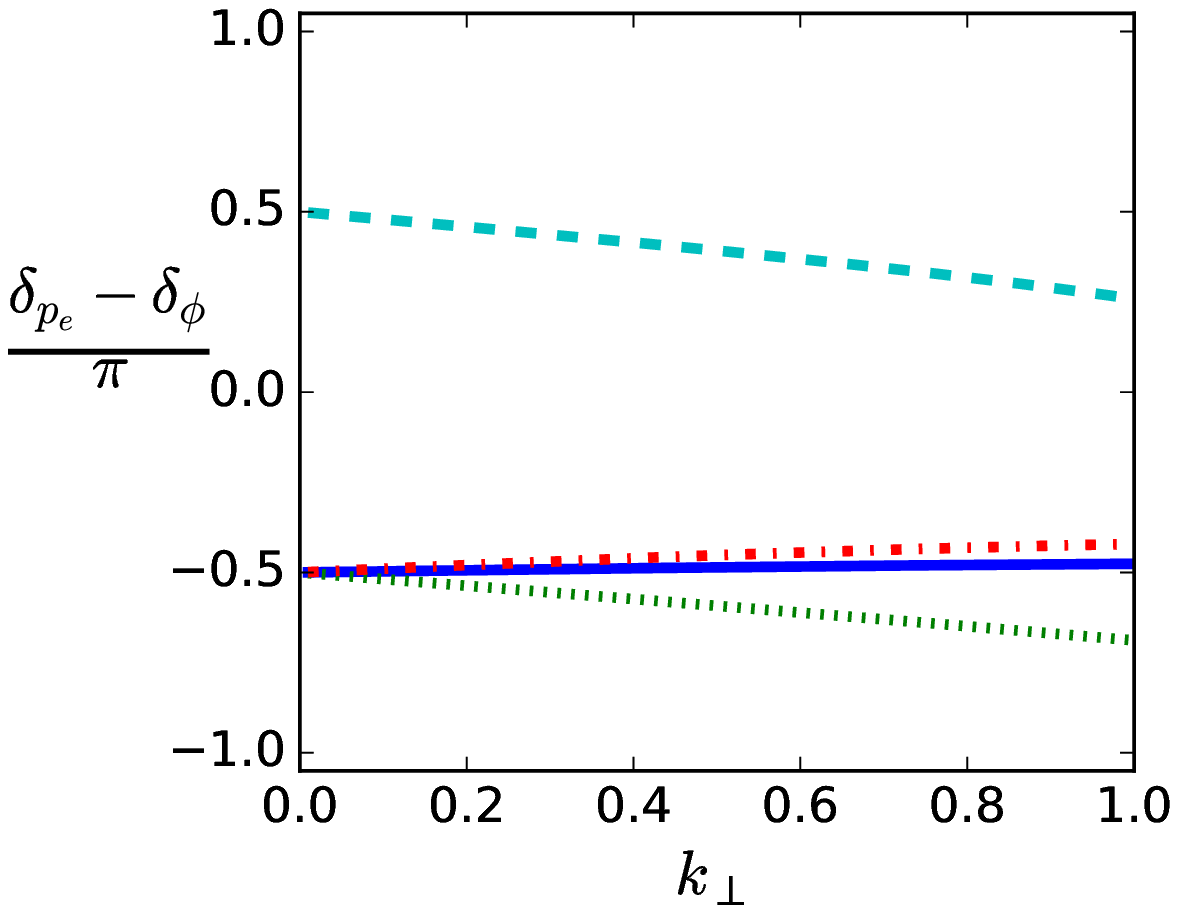} 
 \end{minipage} 
\begin{minipage}[c]{.32\textwidth}
  \par\vspace{0pt}
  \centering 
\includegraphics[width=1\textwidth]{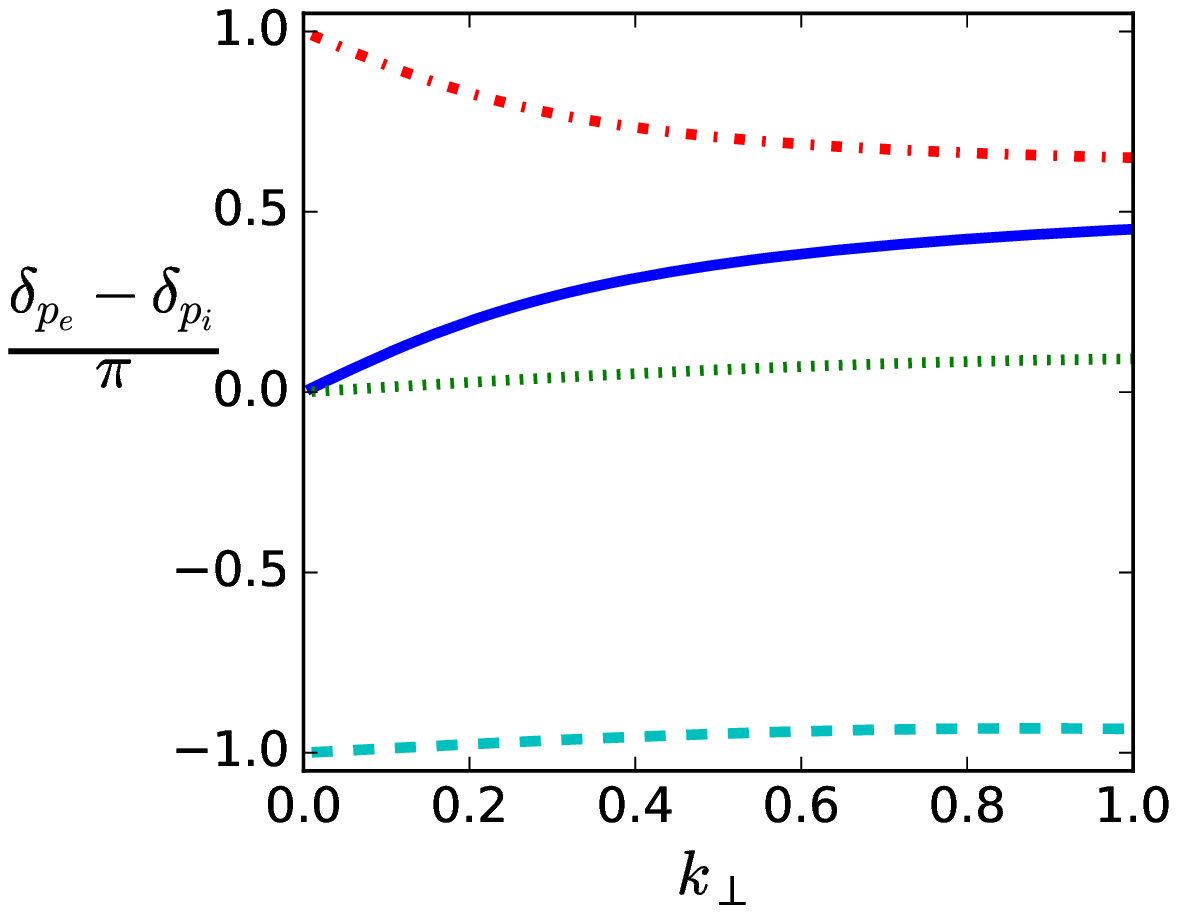} 
 \end{minipage}  
\caption{Linear calculation of phase shift between a) ion pressure and electric potential fluctuations, b) electron pressure and electric potential fluctuations, and c) ion and electron pressure fluctuations as functions of $k_{\perp}$. } 
  \label{fig:QPhase}
\end{figure}

The linear fluctuations are related by
\begin{align}
	\frac{\phi_{\bm{k}}}{p_{i\bm{k}}}
	= \frac{|\phi_{\bm{k}}|}{|p_{i\bm{k}}|}e^{i(\delta_{\phi}- \delta_{p_i})}
	= 
		\frac{3 \lambda (3 \lambda - 2 )}{3\al (3 \bar{\kappa}_i -2)+2  (3 \bar{\kappa}_e-2 )},
		\label{eq:QLPhiPi}\\
	\frac{p_{e\bm{k}}}{p_{i\bm{k}}} 
	= \frac{|p_{e\bm{k}}|}{|p_{i\bm{k}}|}e^{i(\delta_{p_e}- \delta_{p_i})}
	= \frac{(3 \bar{\kappa}_e-2 )(3 \al-2)}{3\al (3 \bar{\kappa}_i -2) + 2 (3 \bar{\kappa}_e-2)}.
\label{eq:QLPePi}
\end{align}
From these expressions the corresponding phase shifts can be calculated (see Fig.~\ref{fig:QPhase}).
As expected the phase shifts between pressure and electric potential fluctuations plotted in Figs.~\ref{fig:QPhase}a and \ref{fig:QPhase}b show that the interchange drive term in Eq.~\reff{eq:fluc_energy} according to Eq.~\reff{eq:FETransSpec} transforms thermal energy into fluctuating energy when the waves are unstable, see Fig.~\ref{fig:DispRel}. We also observe that in the cases where the inverse profile gradient length scales $\bar{\kappa_e} = 1/3$ (cyan) and $\bar{\kappa_i} = 1/3$ (red) are below unity, the direction of the energy flux is reversed even though the waves are unstable. 

\begin{figure}[hbtp]
 \centering
\begin{minipage}[c]{.49\textwidth}
  \par\vspace{0pt}
  \centering 
\includegraphics[width=1\textwidth]{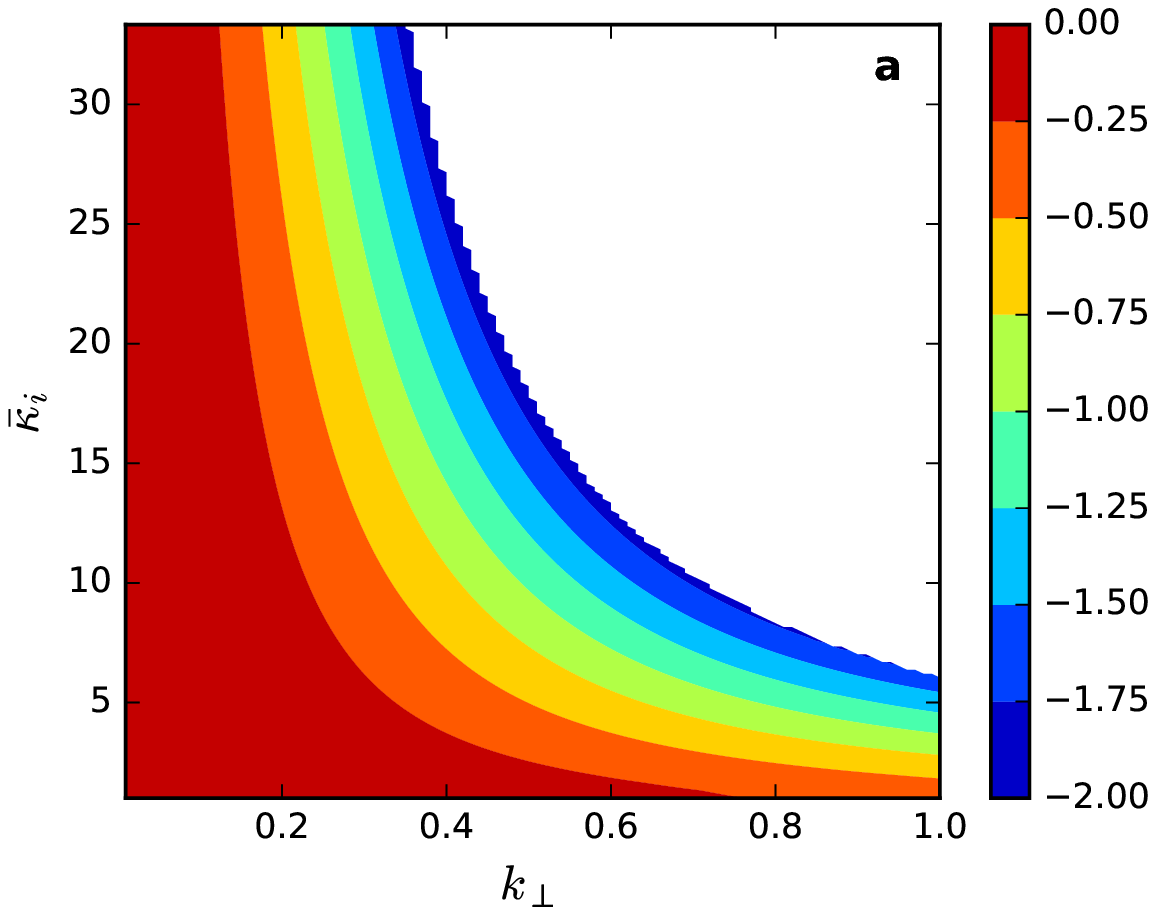}
 \end{minipage}
\begin{minipage}[c]{.49\textwidth}
  \par\vspace{0pt}
  \centering 
\includegraphics[width=1\textwidth]{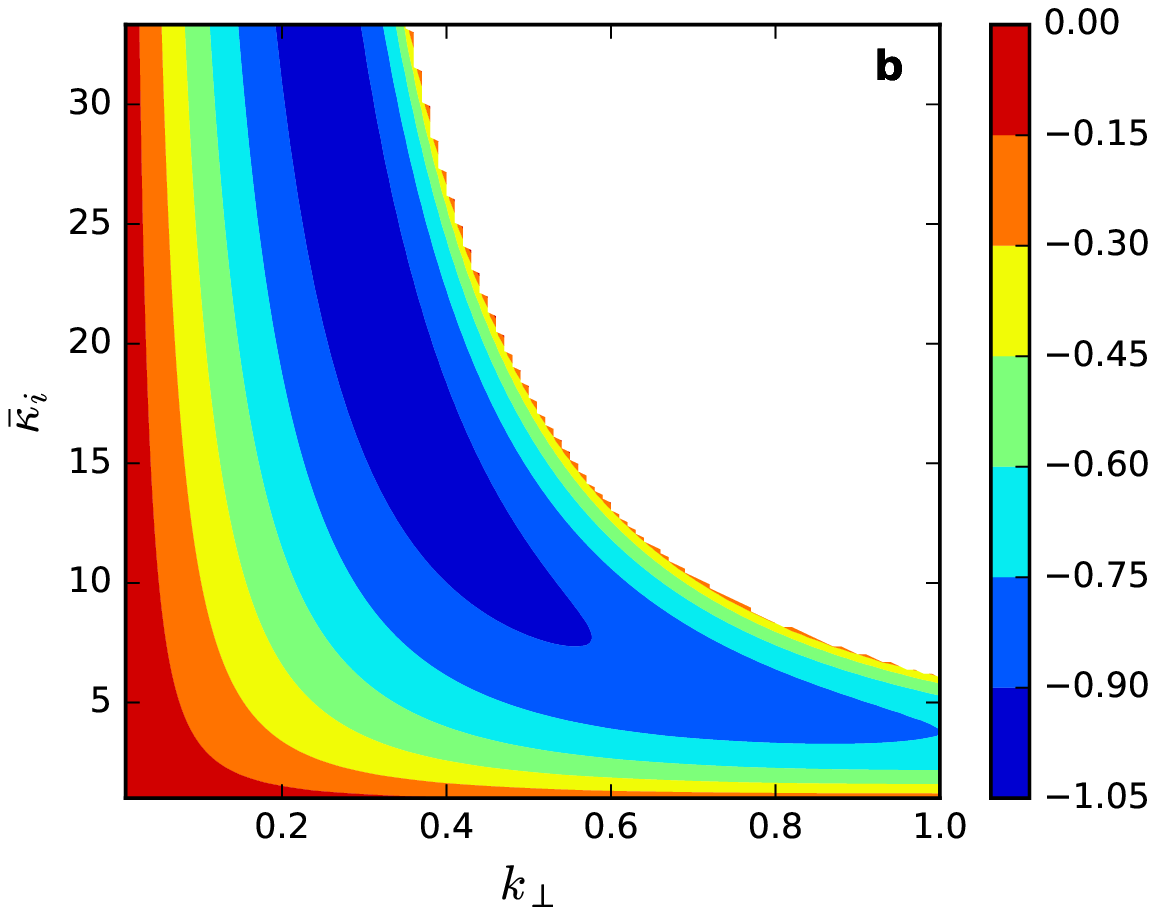} 
 \end{minipage} \\
\begin{minipage}[c]{.49\textwidth}
  \par\vspace{0pt}
  \centering 
\includegraphics[width=1\textwidth]{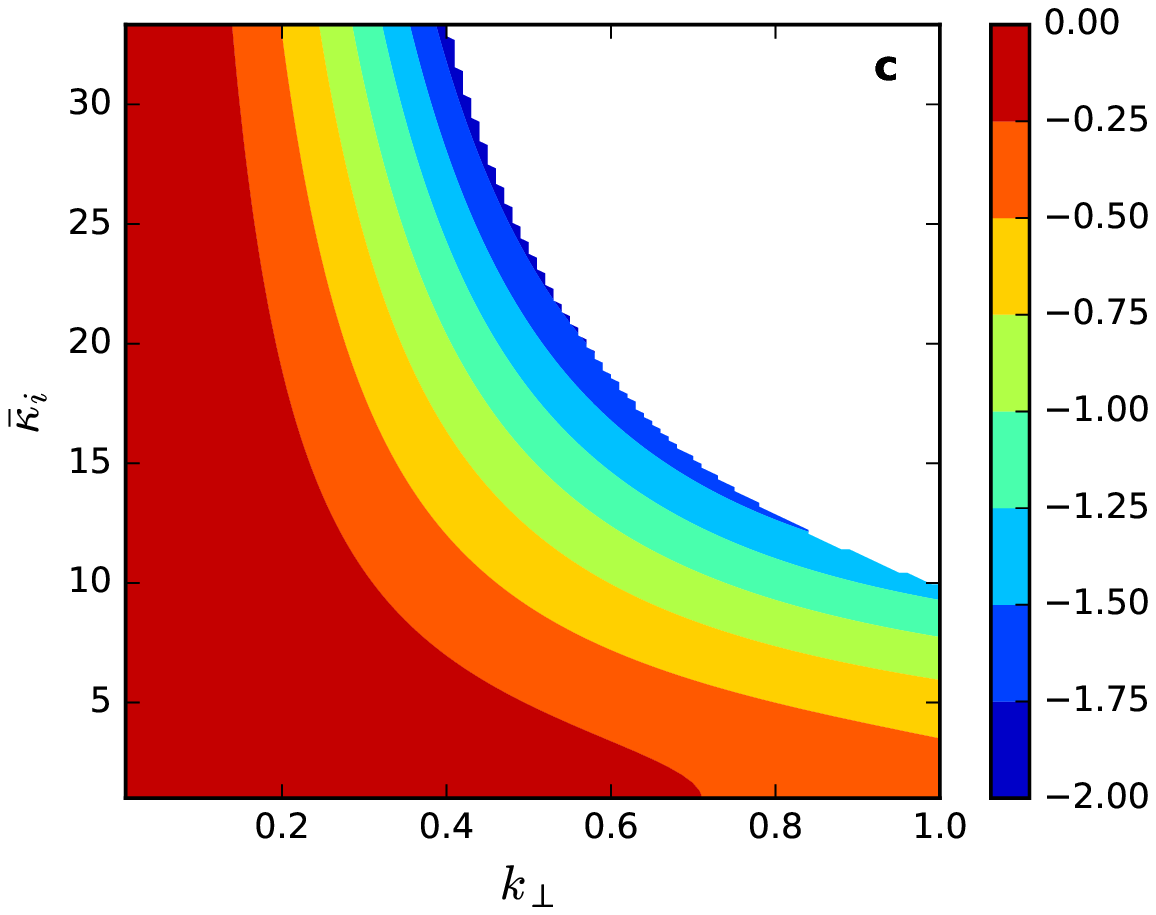}
 \end{minipage}
\begin{minipage}[c]{.49\textwidth}
  \par\vspace{0pt}
  \centering 
\includegraphics[width=1\textwidth]{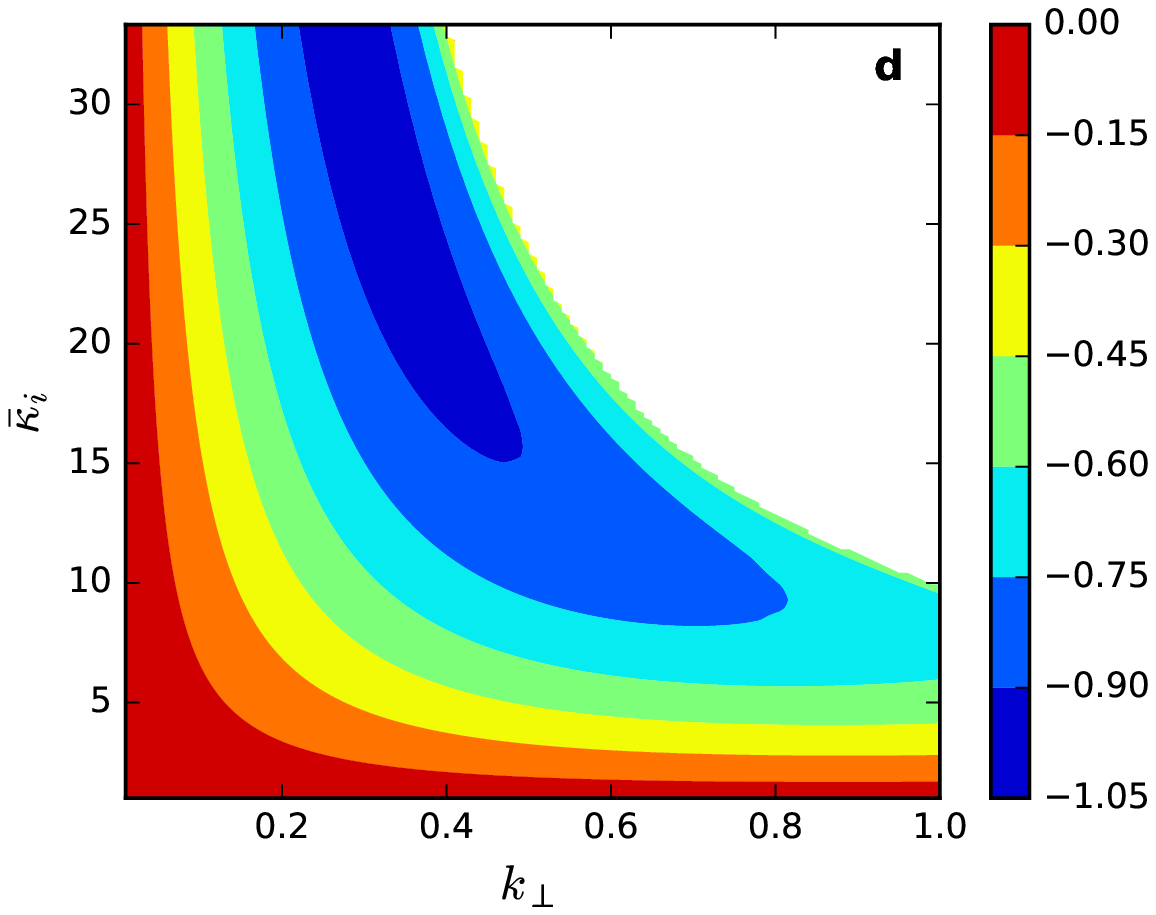} 
 \end{minipage}
 \caption{Comparison of diamagnetic and standard Reynolds stress. When the absolute value  in (\textbf{a}) and (\textbf{c}) is above unity, the first term in the magnitude of diamagnetic Reynolds stress given in Eq.~\reff{eq:QLdiamagneticReynolds} exceeds the standard Reynolds stress. In the white regions (upper right corner in all plots) the solutions are stable. Specifically, the plots show quasi-linear calculations of $\RE\bigg[\frac{p_{i\bm{k}}}{\phi_{\bm{k}}}\bigg]$  for (\textbf{a}) $\bar{\kappa}_e = 1$  and (\textbf{c}) $\bar{\kappa}_e = 10$, and $\IM\bigg[\frac{p_{i\bm{k}}}{\phi_{\bm{k}}}\bigg]$ for (\textbf{b}) $\bar{\kappa}_e = 1$  and  (\textbf{d}) $\bar{\kappa}_e = 10$.  } 
 \label{fig:DiaRey}
\end{figure}

For the analysis of the diamagnetic Reynolds stress given in Eq.~\reff{eq:DiaReynoldsStress}, we employ the quasi-linear approximation. By expressing the ion pressure fluctuations in terms of the potential fluctuations, we get
\begin{align}
	\GA{u_y \pfrac{}{y}p_i}
	= -2 \sum_{k_y>0} k_y \bigg(|\phi_{k_y}|^2 \delta_{\phi}' \RE\bigg[\frac{p_{ik_y}}{\phi_{k_y}}\bigg] 
		+ \frac{1}{2} (|\phi_{k_y}|^2)' \IM\bigg[\frac{p_{ik_y}}{\phi_{k_y}}\bigg]	\bigg).
	\label{eq:QLdiamagneticReynolds}
\end{align}
The first term (see Eq.~\reff{eq:Reynolds_spectral}) equals the Reynolds stress times the real part of the ratio of the ion pressure to the potential. The magnitude of the first term in the diamagnetic Reynolds stress relative to the standard Reynolds stress is therefore simply given by the magnitude  of $\RE[p_{ik_y}/\phi_{k_y}]$. In the quasi-linear treatment this factor can be calculated using Eq.~\reff{eq:QLPhiPi} employing the solution given in Eq.~\reff{eq:sol_disp_rel}. When the absolute value of $\RE[p_{i\bm{k}}/\phi_{\bm{k}}]$ exceeds unity, the first term in the diamagnetic Reynolds stress exceeds the standard Reynolds stress and equivalently the diamagnetic Reynolds stress production term dominates. Quasi-linear calculations of $\RE[p_{i\bm{k}}/\phi_{\bm{k}}]$ as a function of $k_{\perp}$ and $\kappa_i$ are shown in Figs.~\ref{fig:DiaRey}a and \ref{fig:DiaRey}c for $\bar{\kappa}_e = 1$ and $\bar{\kappa}_e = 10$, respectively.
In both cases the diamagnetic Reynolds stress only attains significant values relative to the standard Reynolds stress at intermediate values of $k_{\perp}$ and increase with $\bar{\kappa_i}$. This behavior is expected since the diamagnetic Reynolds stress is an FLR effect, which is expected to become more important as wavelengths and gradient length scales approach ion gyroradius length scales. Steepening of the background electron pressure gradient $\bar{\kappa}_e$ decreases the diamagnetic Reynolds stress relative to the standard Reynolds stress.

The magnitude of the second term of the diamagnetic Reynolds stress given in Eq.~\reff{eq:QLdiamagneticReynolds} depends on the radial gradient of the fluctuating kinetic energy and is therefore only able to drive or damp the mean flow if the fluctuating kinetic energy is radially inhomogenous $\frac{1}{2} (|\phi_{\bm{k}}|^2)' \neq 0$. The magnitude of the fluctuating kinetic energy is not readily accessible through quasi-linear calculations and must be obtained via non-linear numerical calculations. However, the fluctuating energy is multiplied by $\IM\bigg[\frac{p_{i\bm{k}}}{\phi_{\bm{k}}}\bigg]$, and hence regardless of the radial structure of the fluctuating kinetic energy this must be finite for this part of the diamagnetic Reynolds stress to play a role. Quasi-linear calculations of $\IM\bigg[\frac{p_{i\bm{k}}}{\phi_{\bm{k}}}\bigg]$ for $\bar{\kappa}_e = 1$ and $\bar{\kappa}_e = 10$ are shown in Fig.~\ref{fig:DiaRey}b and d, respectively.  In both cases the magnitude is small for $k_{\perp}<0.1$ for all values of $\bar{\kappa}_i$. For $k_{\perp}>0.1$, $\IM\bigg[\frac{p_{i\bm{k}}}{\phi_{\bm{k}}}\bigg]$ is of order unity for most values of $\bar{\kappa}_i$ in the unstable region.  

These calculations should be interpreted with caution for $k_{\perp} \gtrsim 0.5$ because the model is not valid here unless $T_i \ll T_e$. Calculations for wavelengths comparable to the ion gyroradius can be calculated using gyrofluid or gyrokinetic theory. Nonetheless, the calculations show that the diamagnetic Reynolds stress can be important in regions with steep background ion pressure gradients such as in the edge plasma or in internal transport barriers.

Finally, we consider the terms marked "\textbf{c}" and "\textbf{d}" in the mean flow equation \reff{eq:meanflow} and the corresponding terms marked "\textbf{C}","\textbf{D}", and "\textbf{H}" in the energy integrals \reff{eq:meanflow_energy}-\reff{eq:crossterm_energy}. The spectral representations given in Eqs.~\reff{eq:DETransSpec}-\reff{eq:pe_ddy_pi} show that finite contributions by these terms require that the sines of the phase shifts between ion pressure and electric potential as well as between ion and electron pressure fluctuations are finite. Figures~\ref{fig:QPhase}a and \ref{fig:QPhase}c  show that, according to linear theory, these terms yield finite contributions for a wide range of parameters. This observation entails that these mechanisms must be taken into account in the description of mean flows. Specifically, the linear results shown in Fig.~\ref{fig:QPhase}c reveal that the energy transfer term "\textbf{H}", between the ion thermal energy and the residual energy, for most parameters yields an energy transfer from the residual to the ion thermal energy except when electron pressure profiles are nearly flat. The quasi-linear analysis does therefore not indicate the existence of an energy flux from the ion thermal energy via the residual energy to the mean flow energy which bypasses the fluctuating kinetic energy. 

\section{Discussion and conclusions }\label{sec:conclusion}
In this paper we have investigated the influence of ion temperature fluctuations on azimuthal \ExB mean flows in two-dimensional, electrostatic, interchange driven convection. Mean flows perpendicular to the magnetic field are, to leading order,  composed of \ExB and diamagnetic parts. Since the capability of the diamagnetic drift to transport plasma over macroscopic distances is inferior compared to the \ExB drift, only the strength and shear of the \ExB mean flow determines the ability of perpendicular mean flows to suppress turbulence in transport barriers. Our investigations show that in the presence of ion pressure fluctuations there are mechanisms beyond the standard perpendicular \ExB Reynolds stress capable of modifying \ExB mean flows. Specifically, the standard Reynolds stress is accompanied by a diamagnetic Reynolds stress. Quasi-linear analysis indicates that the standard and diamagnetic Reynolds stresses are equally important. In addition to the diamagnetic Reynolds stress we identify two mechanisms capable of modifying \ExB mean flows. Both mechanisms rely on magnetic field inhomogeneity. The first mechanism takes the same form as the interchange energy exchange term, which is responsible for feeding free energy from the free thermal energy into \ExB fluctuations in interchange driven instabilities. This mechanism and the interchange energy exchange term are therefore simultaneously active. The second mechanism relies on phase shifted ion and electron temperature perturbations and is in that respect unique because electric potential fluctuations are not needed. This mechanism provides energy transfer between the ion thermal energy and the mean flow energy completely bypassing electric potential fluctuations. However, quasi-linear analysis shows that the direction of the energy flux inhibits mean flows for most parameters.  

The principal result of this paper is to demonstrate that ion pressure fluctuations also contribute to the generation and sustainment of \ExB mean flows. These additional mechanisms are included in gyrofluid and gyrokinetic models, but are hidden in their mathematical formulation. Only by considering these additional mechanisms explicitly, we will be able to understand \ExB mean flow dynamics and compare our findings with experiment where \ExB mean flows are key ingredients in transport barriers. 

Our analysis was carried out in a simplified two-dimensional drift fluid model describing interchange driven turbulence in the absence of dynamics parallel to the magnetic field. Naturally our results cannot readily be generalized to a toroidal configuration where parallel dynamics plays an important role. In such a more realistic setting several known mechanisms\cite{Scott2010} such as the perpendicular/parallel Reynolds stress, the magnetic flutter contribution, and the Maxwell stress can couple turbulence and mean flows, but we are also convinced that new mean flow mechanisms similar to those presented here exist. It is therefore evident that e.g. the phase shifts between the ion and electron pressures and the electric potential fluctuations will change and that the quasi-linear results presented here will be altered. Nonetheless, the mechanisms for driving \ExB mean flows derived in this paper will persist in a more complete description. Therefore, our analysis points out that the paradigm of Reynolds stress driven mean flows is incomplete and must be supplemented by other mechanisms apparently equivalently capable of modifying \ExB mean flows. 

The existence of mechanisms beyond the Reynolds stress capable of driving \ExB mean flows, could provide an adequate explanation for the contradictory findings in experiments trying to estimate the importance of turbulence driven mean flows \cite{Schmitz2012, Manz2012, Kobayashi2013, Tynan2013, Schmitz2017}. All previous experimental investigations do only account for the pure \ExB Reynolds stress. Other mechanisms for \ExB mean flow generation, including the ones derived here, are not considered in the experiments, but they must be accounted for (or proven negligible in a more complete model) in order to settle the ongoing discussion on the role of the turbulence driven mean flows in transport barriers.

\acknowledgments
This work has been carried out within the framework of the EUROfusion Consortium and has received funding from the Euratom research and training programme 2014-2018 under grant agreement No. 633053. The views and opinions expressed herein do not necessarily reflect those of the European Commission.

\appendix

\section{Approximations to the pressure equations }\label{app:second_order_p_terms}
In this appendix we describe the approximations made in the electron and ion pressure equations \reff{eq:df_pe} and \reff{eq:df_pi} that lead to the reduced model given in Eqs.~\reff{eq:w_norm}-\reff{eq:elecpressure_norm}.

First, in both pressure equations \reff{eq:df_pi} and \reff{eq:df_pe}, we neglect the order $\epsilon^2$ compressional contributions $3/2 p_a \nabla \cdot \bm{u}_E$ in the $3/2\nabla \cdot (p_a \bm{u}_E)$ terms. The advection parts are evaluated with a constant magnetic field magnitude: $3/2 \bm{u}_E  \cdot \nabla p_a \simeq  3/2\frac{B}{B_0} \bm{u}_E \cdot \nabla p_a$. This approximation leaves the energy theorem unchanged, and shown in Sec.~\ref{sec:energythm}, the exchange between \ExB energy and the thermal reservoirs is mediated\cite{pop2003BDS} by the \ExB compression terms $p_a \nabla \cdot \bm{u}_E$. These energy exchange terms are therefore retained. 

Advection of pressure by the diamagnetic drift vanishes due to the "diamagnetic cancellation": 
\begin{align}
\frac{3}{2}\nabla \cdot(p_a \bm{u}_{Da})
+p_a\nabla \cdot\bm{u}_{Da}
+ \nabla \cdot \bm{q}^*_{\perp a} 
= \frac{5}{2}\curl\big(\frac{\bhat}{q_aB}\big) \cdot \nabla (p_a T_a)
\end{align}
The curvature term on the right hand side is of order $\epsilon^2$ and since it does not influence the conservation of energy we neglect all diamagnetic drift terms in the pressure equations. 

Lastly, all terms including the polarization and gyroviscous drifts in the ion pressure equation are of order $\epsilon^2$. Again, we neglect the divergence terms $\frac{3}{2}\nabla \cdot \big(p_i[\bm{u}_{p_i}+ \bm{u}_{\pi_i} ]\big)$ as they have no influence on the energy theorem, and we keep the compressional contributions $p_i\nabla \cdot (\bm{u}_{p_i}+ \bm{u}_{\pi_i})$ which permits energy exchange between the ion thermal energy and \ExB kinetic energy\cite{pop2003BDS}. The thin-layer approximation must also made here (see Eq.~\reff{eq:thinlayer_up}) in order to conserve energy. In other words, we must make the same approximations to the first order drifts in all equations\cite{madsen2016}. This requirement is also necessary for establishing the correspondence between drift fluid and gyrofluid models\cite{scott:102318}. The resulting pressure equations used for the studies in this paper are given in Eqs.~\reff{eq:w_norm}-\reff{eq:elecpressure_norm}.

\section{Energy conservation mean flows in full 2D interchange model}\label{sec:appendix} 
In the following we describe the implications of retaining all the second order terms in the pressure equations which were considered in appendix \ref{app:second_order_p_terms}. Specifically, we show that: a) the mean flow equation \reff{eq:meanflow} is unaltered, b) the energy theorem Eq.~\reff{eq:energy_theorem} and the energy transfer channels in Eqs.~\reff{eq:d_dt_drift_energy}-\reff{eq:d_dt_elec_energy} are the same, and c) all energy transfer channels in the \ExB mean flow energy equation \reff{eq:meanflow_energy} remain, but with modified prefactors. Furthermore, an additional transfer term due to advection of ion temperature by the ion polarization drift is added. 

When all second order terms are retained in the electron and ion pressure equations the 2D interchange model in slab geometry and Gyro-Bohm normalized units (see Eq.~\reff{eq:gyro_bohm_normalization}) is given as:
\begin{subequations}
	\begin{align}
	\frac{D}{Dt}n
	-  n \xi \pfrac{}{y} \phi 
	+  \xi \pfrac{}{y}p_e
	&= \Lambda_n
	\label{eq:n_full}\\
	\nabla \cdot \big(\frac{d}{dt}  \np \phi^*\big) 
	+\xi\pfrac{}{y}(p_e + p_i)    
	&= \Lambda_w,\label{eq:w_full}
	\\
	\frac{3}{2}\frac{D}{Dt}p_i
	-\frac{5}{2} p_i \xi\pfrac{\phi}{y}
	- \frac{5}{2} \xi \pfrac{}{y}\frac{p_i^2}{n}
	+  \frac{5}{2}p_i \xi \pfrac{}{y}(p_e + p_i)    
	- \frac{3}{2} \bigg[\frac{d}{dt} \nabla \phi^* \bigg]\cdot \nabla p_i 
	&= \Lambda_{p_i},           
	\label{eq:ionpressure_full}\\
	\frac{3}{2}\frac{D}{Dt}p_e
	- p_e \xi\pfrac{\phi}{y}
	+ \frac{5}{2} \xi \pfrac{}{y}\frac{p_e^2}{n}
	&= \Lambda_{p_e},           
	\label{eq:elecpressure_full}
	\end{align}
\end{subequations}
where we introduce the material derivative with non-constant magnetic field 
\begin{align}
	\frac{D}{Dt} = \pfrac{}{t} +  \frac{1}{B(x)}\{\phi, ~\}. 
\end{align}
The varying magnetic field is dictated by energy conservation. The diamagnetic pressure and heat fluxes in the ion and electron pressure equations are the only non-collisional terms which explicitly depend on the particle density $n$. Retention of these higher order terms demands that the particle density equation is added to the model. Note that in comparison to the applied model Eq.~\reff{eq:elecpressure_norm}-\reff{eq:ionpressure_thinlayer}, the prefactors on the $\xi$-dependent terms are altered. Furthermore, the full model also includes the advection of ion pressure by the ion polarization drift; last term on the right hand side of \reff{eq:ionpressure_full}. This term is neglected in existing drift fluid models\cite{RogersPRL1998, Scott2005NJP,Halpern2016388}.

\paragraph{Mean flow equation}
The \ExB mean flow equation \reff{eq:meanflow} is derived from the vorticity equation \reff{eq:w_norm}. Since the vorticity equation is not changed, nor is the \ExB mean flow equation.

\paragraph{Energy theorem}
The energy theorem Eq.~\reff{eq:energy_theorem} is also left unchanged. The theorem is derived by integrating: i) the pressure equations \reff{eq:ionpressure_full}-\reff{eq:elecpressure_full} and ii)"$- \phi$" times the vorticity equation \reff{eq:w_full}, over the domain. Summation of the integrals yield the desired result. 

The electron pressure integral is only modified by the diamagnetic term (last term on the left hand side of Eq.~\reff{eq:elecpressure_full}). Since the slab geometry is periodic in the y-direction this term trivially vanishes when integrated over the 2D domain. This also holds true for the corresponding term in the ion pressure equation. The remaining new terms in the ion pressure equation \reff{eq:ionpressure_full} also vanish since they form divergence terms which only yield surface terms, which are neglected in the derivation of the energy theorem Eq.~\reff{eq:energy_theorem}. For instance the \ExB terms are combined using
\begin{align}
	\frac{1}{B}\{\phi,p_i\} - p_i \xi \pfrac{}{y}\phi = \nabla \cdot (p_i \bm{u}_E ). 
\end{align}
Similar manipulations of the polarization drift, ion pressure flux term result in a divergence term which only gives a surface contribution to the energy integrals. Therefore, also the energy transfer channels in Eqs.~\reff{eq:d_dt_drift_energy}-\reff{eq:d_dt_elec_energy} remain the same. 

\paragraph{\ExB mean flow energy}
An equation governing the time evolution of kinetic energy $E_0$ associated with the mean \ExB flow $u_0 = \partial_x \GA{\phi}$, is obtained by integrating the product of $\GA{\phi}$ times the vorticity equation \reff{eq:w_full}:
\begin{align}
\frac{d}{dt} E_0 =
\int d\bm{x} \,  
\bigg[
\underset{\textbf{A}}{ \GA{u_y u_x}}
-\underset{\textbf{B}}{\GA{u_y \pfrac{p_i}{y} }}
-\underset{\textbf{C}^*	}{\frac{5}{3}\xi \GA{p_i \pfrac{p_e}{y}} }
-\underset{\textbf{D}^*}{\frac{5}{3}\xi\GA{ p_i  u_x}}
+\underset{\textbf{B}^*}{\xi x \pfrac{}{x}(\phi \pfrac{}{y} p_i) }
-\underset{\textbf{D}^\dagger}{\nabla p_i \cdot \frac{d}{dt} \nabla \phi^*  }
\bigg] u_0' \notag \\
- \GA{\phi} \bigg[
\underset{\textbf{E}}{\GA{\Lambda_w}    
	-\frac{2}{3} \pfrac{^2}{x^2}\GA{\Lambda_{p_i}}}
\bigg].
\label{eq:meanflow_energy_full}
\end{align}
In comparison with the \ExB mean flow energy theorem Eq.~\reff{eq:meanflow_energy}, the standard \ExB Reynolds stress and the diamagnetic Reynolds stress, "\textbf{A}" and "\textbf{B}", respectively, are left unchanged. The terms "$\textbf{C}^*$" and "$\textbf{D}^*$" have the same from as in the original equation \reff{eq:meanflow_energy} but the coefficients are changed. The "$\textbf{B}^*$" is new. It appears because the \ExB drift entering the vorticity equation \reff{eq:w_full} is evaluated with a constant magnetic field whereas the magnetic field in the ion pressure equation \reff{eq:ionpressure_full} is x-dependent $B^{-1} = B_0^{-1} (1 + \xi x)$. This discrepancy is an inherent consequence of the thin-layer approximation in drift fluid models\cite{madsen2016,pop2003BDS}. The term is $\epsilon_B$ smaller than the leading order terms, see Eq.~\reff{eq:driftordering}. Lastly, an energy transfer channel "$\textbf{D}^\dagger$" appears. This additional energy transfer mechanism is due to the advection of ion pressure by the ion polarization drift. The drift ordering presumes that the polarization drift is small compared with the \ExB and diamagnetic drifts and hence this additional energy transfer term is presumed small compared to e.g. the diamagnetic Reynolds stress "\textbf{B}". Lastly, we note that the inclusion of the diamagnetic terms in the pressure equations do not give rise to new energy transfer channels as expected. 

In conclusion, the principal result of this paper is that there are non-negligible mechanisms beyond the standard  \ExB Reynolds stress which modify the \ExB mean flow. The neglect of higher order terms in the pressure equations do not alter this result, the inclusion of these terms, on the other hand, complicate the derivations and the analysis.


\begin{thebibliography}{10}
\bibitem{Alfven1940}
H.~Alfv{\'e}n,
\newblock {\em Arkiv Mat. Astr. Fysik}, 27, 1 (1940).

\bibitem{Belova_2001}
E.~V.~Belova,
\newblock {\em Phys. Plasmas}, 8, 3936 (2001).

\bibitem{Biglari1990}
H.~Biglari, P.~H.~Diamond, and P.~W.~Terry,
\newblock {\em Phys. Fluids B}, 2, 1 (1990).

\bibitem{Brizard_1992}
A.~Brizard,
\newblock {\em {Phys. Fluids B}}, 4, 1213 (1992).

\bibitem{brizard_Hahm_review2007}
A.~J.~Brizard and T.~S.~Hahm,
\newblock {\em Rev. Mod. Phys.}, 79, 421 (2007).

\bibitem{Brizard2011}
A.~J.~Brizard and N.~Tronko,
\newblock {\em Phys. Plasmas}, 18, 082307 (2011).

\bibitem{Burrell1992}
K.~H.~Burrell , T.~N.~Carlstrom, E.~J.~Doyle, D.~Finkenthal, P.~Gohil, R.~J.~Groebner, D.~L.~Hillis, J.~Kim, H.~Matsumoto, R.~A.~Moyer, T.~H.~Osborne, C.~L.~Rettig, W.~A.~Peebles,
  T.~L.~Rhodes, H.~StJohn, R.~D.~Stambaugh, M.~R.~Wade, and J.~G.~Watkins,
\newblock {\em Plasma Phys. Controlled Fusion}, 34, 1859 (1992).

\bibitem{Connor2000}
J.~W.~Connor and H.~R.~Wilson,
\newblock {\em Plasma Phys. Controlled Fusion}, 42, R1 (2000).

\bibitem{Diamond2005}
P~H.~Diamond, S-I.~Itoh, K.~Itoh, and T.~S.~Hahm,
\newblock {\em Plasma Phys. Controlled Fusion}, 47, R35 (2005).

\bibitem{Diamond1991}
P.~H.~Diamond and Y.-B.~Kim,
\newblock {\em Phys. Fluids B}, 3, 1626 (1991).

\bibitem{Dif-Pradalier2009}
G.~Dif-Pradalier, V.~Grandgirard, Y.~Sarazin, X.~Garbet, and Ph.~Ghendrih,
\newblock {\em Phys. Rev. Lett.}, 103, 065002 (2009).

\bibitem{Garcia_diamagnetic_fluid_part}
O.~E.~Garcia,
\newblock {\em European J. Phys.}, 24, 331 (2003).

\bibitem{Garcia2003PRE}
O.~E.~Garcia and N.~H.~Bian,
\newblock {\em Phys. Rev. E}, 68, 047301 (2003).

\bibitem{garcia_inter}
O.~E.~Garcia, N.~H.~Bian, V.~Naulin, A.~H.~Nielsen, and J.~Juul~Rasmussen.
\newblock {\em Physica Scripta}, T122, 104 (2006).

\bibitem{Garcia_POP_062309}
O.~E.~Garcia, V.~Naulin, A.~H.~Nielsen, and J.~Juul~Rasmussen.
\newblock {\em {Phys. Plasmas}}, 12, 62309 (2005).

\bibitem{Halpern2016388}
F.~D.~Halpern, P.~Ricci, S.~Jolliet, J.~Loizu, J.~Morales, A.~Mosetto, F.~Musil,
F.~Riva, T.~M.~Tran, and C.~Wersal,
\newblock {\em J. Comp. Phys.}, 315, 388 (2016).


\bibitem{Hinton_Horton_1971}
F.~L.~Hinton and C.~W.~Horton,
\newblock {\em Phys. Fluids}, 14, 116 (1971).

\bibitem{knorr_1988}
G.~Knorr, F.~R.~Hansen, J.~P.~Lynov, H.~L.~P{\'e}cseli, and J.~Juul~Rasmussen,
\newblock {\em Physica Scripta}, 38, 829 (1988).

\bibitem{Kobayashi2013}
T.~Kobayashi, K.~Itoh, T.~Ido, K.~Kamiya, S.-I. Itoh, Y.~Miura, Y.~Nagashima,
A.~Fujisawa, S.~Inagaki, K.~Ida, and K.~Hoshino,
\newblock {\em Phys. Rev. Lett.}, 111, 035002 (2013).

\bibitem{Krommes2013}
J.~A.~Krommes and G.~W.~Hammett.
\newblock {Report of the Study Group GK2 on Momentum Transport in
  Gyrokinetics}.
\newblock {\em PPPL Technical Report {\#}4945}, (2013).

\bibitem{kundu2010fluid}
P.~K.~Kundu and I.~M.~Cohen.
\newblock {\em {Fluid Mechanics}}.
\newblock Elsevier Science, 2010.

\bibitem{madsen2016}
J.~Madsen, V.~Naulin, A.~H.~Nielsen, and J.~Juul Rasmussen,
\newblock {\em Phys. Plasmas}, 23, 032306 (2016).

\bibitem{MadsenPPCF2015}
J.~Madsen, J.~Juul~Rasmussen, V.~Naulin, A.~H.~Nielsen, and F.~Treue,
\newblock {\em Plasma Phys. Controlled Fusion}, 57, 54016 (2015).

\bibitem{madsen2013GF}
J.~Madsen,
\newblock {\em Phys. Plasmas}, 20, 072301 (2013).

\bibitem{jmad2011FLRBlob}
J.~Madsen, O.~E.~Garcia, J.~S.~Larsen, V.~Naulin, A.~H.~Nielsen, and J.~Juul  Rasmussen,
\newblock {\em Phys. Plasmas}, 18, 112504 (2011).

\bibitem{Manz2012}
P.~Manz, G.~S. Xu, B.~N. Wan, H.~Q. Wang, H.~Y. Guo, I.~Cziegler, N.~Fedorczak,
C.~Holland, S.~H. M{\"u}ller, S.~C. Thakur, M.~Xu, K.~Miki, P.~H. Diamond,
and G.~R. Tynan,
\newblock {\em Phys. Plasmas}, 19, 072311 (2012).


\bibitem{Nielsen2015}
A.~H.~Nielsen, G.~S.~Xu, J.~Madsen, V.~Naulin, J.~Juul~Rasmussen, and B.~N.~Wan,
\newblock {\em Phys. Lett. A}, 379, 3097 (2015).


\bibitem{Parra2010}
F.~I. Parra and P.~J. Catto,
\newblock {\em Plasma Phys. Controlled Fusion}, 52, 045004 (2010).


\bibitem{Ramos2005}
J.~J.~Ramos,
\newblock {\em Phys. Plasmas}, 12, 1 (2005).

\bibitem{rasmussen_EPS_2015}
J~Juul~Rasmussen, A.~H.~Nielsen, J.~Madsen, V.~Naulin, and G.~S.~Xu
\newblock {\em Plasma Phys. Controlled Fusion}, 58, 14031 (2016).

\bibitem{Reynolds1895}
O.~Reynolds,
\newblock {\em Philosophical Transactions of the Royal Society A: Mathematical,
  Physical and Engineering Sciences}, 186, 123 (1895).


\bibitem{RogersPRL1998}
B.~N. Rogers, J.~F. Drake, and A.~Zeiler,
\newblock {\em Phys. Rev. Lett.}, 81, 4396 (1998).

\bibitem{Schmitz2017}
L.~Schmitz,
\newblock {\em Nuclear Fusion}, 57, 025003 (2017).

\bibitem{Schmitz2012}
L.~Schmitz, L.~Zeng, T.~L. Rhodes, J.~C. Hillesheim, E.~J. Doyle, R.~J.
Groebner, W.~A. Peebles, K.~H. Burrell, and G.~Wang,
\newblock {\em Phys. Rev. Lett.}, 108, 155002(2012).

\bibitem{Scott1998}
B.~Scott,
\newblock {\em Contrib. Plasma Phys.}, 38, 171 (1998).

\bibitem{pop2003BDS}
B.~Scott,
\newblock {\em Phys. Plasmas}, 10, 963 (2003).


\bibitem{Scott200353PLA}
B.~Scott,
\newblock {\em Phys. Lett.}, 320, 53 (2003).

\bibitem{Scott2010}
B.~Scott and J.~Smirnov,
\newblock {\em Phys. Plasmas}, 17, 112302 (2010).

\bibitem{Scott2005NJP}
B.~D.~Scott,
\newblock {\em New J. Phys.}, 7, 92 (2005).

\bibitem{Scott_2005}
B.~D.~Scott,
\newblock {\em Phys. Plasmas}, 12, 102307 (2005).

\bibitem{scott:102318}
B.~D.~Scott,
\newblock {\em Phys. Plasmas}, 14 ,102318 (2007).

\bibitem{Smolyakov1998}
A.~I.~Smolyakov,
\newblock {\em Can. J. Phys}, 76, 321 (1998).

\bibitem{Tsai1970}
S.~Tsai,
\newblock {\em Phys. Fluids}, 13, 2108 (1970).

\bibitem{Tynan2013}
G.~R.~Tynan et al,
\newblock {\em Nuclear Fusion}, 53, 073053 (2013).

\bibitem{wagner_2007}
F.~Wagner,
\newblock {\em Plasma Phys. Controlled Fusion}, 49, B1(2007).

\bibitem{weiland1999collective}
J.~Weiland,
\newblock {\em {Collective Modes in Inhomogeneous Plasmas: Kinetic and Advanced
  Fluid Theory}},
\newblock {Series in Plasma Physics and Fluid Dynamics}. Taylor \& Francis,
  1999.

\bibitem{wiesenberger2014}
M.~Wiesenberger, J.~Madsen, and A.~Kendl,
\newblock {\em Phys. Plasmas}, 21, 092301 (2014).

\bibitem{Zeiler}
A.~Zeiler, J.~F.~Drake, and B.~Rogers,
\newblock {\em {Phys. Plasmas}}, 4, 2134 (1997).

\end{thebibliography}

\end{document}